\documentclass[11pt]{article}
\usepackage{jheppub}
\usepackage{amsmath,amssymb,amsfonts,amsthm,graphicx,physics}
\usepackage{setspace}
\usepackage{subcaption}
\usepackage{dsdshorthand}
\usepackage{physics}
\usepackage{leftindex}
\usepackage{bbm}

\usepackage[left=1in,right=1in,top=1in,bottom=1.2in]{geometry}
\usepackage[dvipsnames]{xcolor}
\usepackage{simpler-wick}
\usepackage{tensor}

\usepackage{tikz}
    \usepackage{amssymb,amsfonts,amsmath}
    \usepackage{tkz-euclide}
        \usetikzlibrary{arrows,calc,patterns}
\usepackage{pgfplots}

\definecolor{darkblue}{rgb}{0.1,0.1,.7}
\definecolor{purple}{rgb}{0.6,0,0.6}
\definecolor{orange}{rgb}{0.9,0.6,0}
\definecolor{llgray}{rgb}{0.9,0.9,1}
\definecolor{dgreen}{rgb}{0,0.5,0}
\definecolor{outsideyellow}{rgb}{0.96,0.86,0.41}
\definecolor{insideyellow}{rgb}{0.99,0.96,0.82}
\definecolor{observerred}{rgb}{0.93,0.27,0.13}
\definecolor{mattergreen}{rgb}{0,0.63,0.29}
\usepackage[]{latexsym}
\usepackage[utf8]{inputenc}
\usepackage{geometry}
\usepackage{amscd}
\usepackage[all,cmtip]{xy}
\usepackage{mathrsfs}

\usepackage[margin=10pt,font=small,labelfont=bf]{caption}
\usepackage{dsdshorthand}
\usepackage{changepage}
\usepackage{setspace}
\usepackage{tikz}
\usepackage{subcaption}
\usepackage{marvosym}
\usepackage{tensor}
\usepackage{CJKutf8}
\usepackage{empheq}

\title{ 
    The gravitational index of $5d$ black holes and black strings
} 
\author[1]{Jan Boruch,}
\author[2,3]{Roberto Emparan,}
\author[1]{Luca V.~Iliesiu,}
\author[4]{Sameer Murthy}

\def\dd{d}

\renewcommand{\=}{\; = \;}
\newcommand{\nv}{n_\text{v}}
\newcommand{\CK}{\mathcal{K}}

\newcommand{\mK}{\mathcal{K}}

\newcommand{\CN}{\mathcal{N}}
\newcommand{\tads}{t_{\text{BTZ}}}

\newcommand{\bast}{\boldsymbol{\ast}}

\newcommand{\barZ}{\overline{Z}} 
\newcommand{\barOmega}{\overline{\Omega}} 
\newcommand{\IIA}{{\text{IIA}}}

\newcommand{\xvec}{\bold{x}}

\newcommand{\ii}{{\rm i}}

\affiliation[1]{Center for Theoretical Physics and Department of Physics, University of California, Berkeley, California 94720, U.S.A.}

\affiliation[2]{Institució Catalana de Recerca i Estudis Avançats (ICREA),
 Passeig Lluis Companys, 23, 08010 Barcelona, Spain}
\affiliation[3]{Departament de Física Quàntica i Astrofísica and
  Institut de Ciències del Cosmos, 
 Universitat de Barcelona, 08028 Barcelona, Spain}
 
\affiliation[4]{Department of Mathematics, King's College London, The Strand, London WC2R 2LS, UK}

\abstract{
The supersymmetric index of $5d$ black strings and spinning black holes in M-theory is related to that of $4d$ black holes in type IIA supergravity when both theories are compactified on the same  Calabi-Yau threefold. 
We find the finite-temperature saddles for the $5d$ gravitational supersymmetric index 
by uplifting the recently found attractor saddles of the corresponding $4d$ index.
We study uplifts for two types of geometries: $5d$ black holes and $5d$ black strings. For $5d$ black holes, the uplift guarantees that the index of $4d$ and $5d$ black holes match. 
For $5d$ black strings, the saddle reproduces the microscopic index at leading order in $G_N$, even without the conventional decoupling limit taken in AdS/CFT. 
In particular, when the temperature is set to be finite in the $5d$ flat space region, the black string index is computed from an asymptotically flat solution where the AdS throat is absent. 
Further, as the temperature is lowered and eventually becomes infinitesimally small in the flat space region, the 
solution admits a novel decoupling limit in which the AdS$_3$ throat takes the form of a finite-temperature BTZ black hole that is known to compute the index in AdS$_3$/CFT$_2$.  
This represents the first step towards understanding holography for supersymmetric observables in flat space, away from the decoupling limit. 
}
\date{\today}

\begin{document}

\maketitle

\section{Introduction}

The match of the macroscopic entropy of black holes to a microscopic count over states 
is one of the main successes of string theory \cite{Sen:1995in, Strominger:1996sh}. Supersymmetry plays a critical role in this match. 
Only for black hole microstates protected by supersymmetry can one give a reliable microscopic 
count in the form of a supersymmetric index. Such an index is computed at a non-zero temperature 
and counts bosonic and fermionic microstates with an opposite sign. 
The match between this index and the thermodynamic entropy of extremal black hole solutions 
in supergravity has long been taken for granted despite the fact that such solutions only exist 
at zero temperature and their Bekenstein-Hawking entropy counts bosonic and fermionic 
states with the same sign.

Recent work has found the true gravitational saddle-point contributions to the index 
of black holes in AdS \cite{Cabo-Bizet:2018ehj} and in flat space \cite{Iliesiu:2021are, Boruch:2023gfn}.\footnote{Such saddles were also generalized to other asymptotic AdS spaces  in~\cite{Cassani:2019mms,Bobev:2019zmz, Bobev:2020pjk,Larsen:2021wnu,BenettiGenolini:2023rkq} and to string sized black holes in flat space \cite{Chowdhury:2024ngg, Chen:2024gmc}. Such saddles were also used in \cite{Heydeman:2020hhw, Boruch:2022tno, Sen:2023dps, H:2023qko, Anupam:2023yns, Hegde:2024bmb} to calculate non-trivial corrections to the gravitational index, including fluctuations from the Schwarzian mode \cite{Heydeman:2020hhw,Iliesiu:2021are, Boruch:2022tno}. 
}  
Such saddles are defined for non-zero temperatures and have a finite throat in the interior
of the geometry. 
Further, they admit periodic boundary conditions for fermionic fields around the thermal circle 
and, consequently, count bosonic and fermionic states with opposite signs. 
The contribution of these saddles to the index, determined at leading order by their on-shell action, turns out to be independent of the temperature; this is precisely what is expected of an index since the contribution of non-BPS bosonic states, whose energy is unprotected, precisely cancels that of non-protected fermionic states. Surprisingly, this occurs despite most other properties of the 
solution---such as the area of the black hole or the value of the moduli on the horizon---being temperature-dependent.

It is natural to ask the broader question: what are the gravitational saddles corresponding to the 
supersymmetric index of general black objects in string theory? This includes a generalization to higher-dimensional 
black holes and---with a much larger scope---to the plethora of black strings and branes that exist in string theory (see \cite{Mohaupt:2000gc} for a review).  
Indeed, the microscopic index of black branes, captured by CFT$_{d\,>\,1}$, can be said to be under
much better control than the microscopic index of black holes captured by some CFT$_1$ whose details are as yet unknown. 
In this paper, we begin this program and obtain the saddles corresponding to black holes and black strings 
in $5d$ supergravity in a Calabi-Yau compactification of M-theory.

Just like for black holes, we show that the index of black strings in asymptotically flat space can always be computed at 
finite temperatures. The corresponding gravitational saddles no longer have a decoupling limit in which 
an AdS$_{3}$ throat can be separated from the asymptotically flat region. 
Nevertheless, the contribution of such saddles to the gravitational index once again turns out to be 
independent of the temperature. 
For black strings in a $5d$ Calabi-Yau compactification of M-theory, we show that this contribution agrees 
with the leading order behavior of the corresponding microscopic index, namely the elliptic genus of the 
MSW SCFT$_2$ living on the string \cite{Maldacena:1997de}. We emphasize that this agreement holds even though the 
gravitational configurations are non-extremal flat-space geometries where the AdS$_{3}$ throat is absent.  
To be more precise, the elliptic genus of the MSW SCFT$_2$ captures one sector of the index of $5d$ flat space, 
which takes into account more general supersymmetric objects that go beyond the scope of this 
paper.\footnote{As we shall discuss in section \ref{sec:discussion}, other objects that should contribute to the index 
include bound configurations of multiple black strings and black holes and their possible excitations.} 
Nevertheless, this represents a first step towards understanding holography for supersymmetric observables 
in flat space, away from the decoupling limit \cite{Maldacena:1997re}.

 \begin{figure}[t!]
    \centering
    \includegraphics[width=0.99\linewidth]{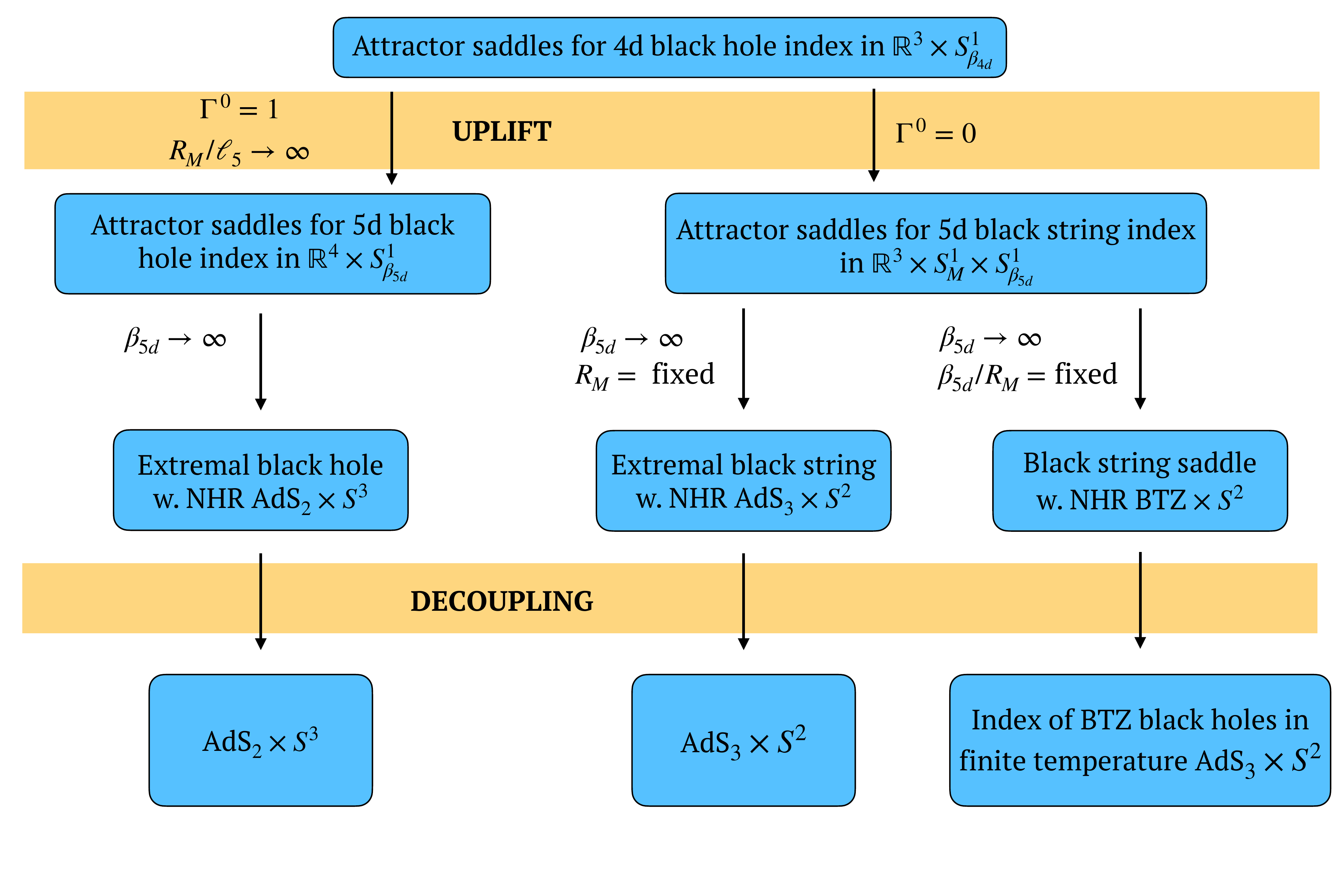}
    \caption{The different uplift solutions that we study, the properties of their low-temperature limits, and the geometries obtained by considering the decoupling limit in the throat of the geometry. $\Gamma^0$ is the D6-brane charge in 4$d$ and NUT charge in 5$d$, so the uplifted solutions with $\Gamma^0=0,1$ are topologically different. $R_M$ is the radius of the Kaluza-Klein five-dimensional (or M-theory) circle.}
    \label{fig:uplift_diagram}
\end{figure}

On a technical level, we construct both the finite-temperature black hole and black string saddles of the $5d$ index 
by uplifting the new $4d$ attractor saddles that capture the index of black holes in $\mathcal N=2$ ungauged supergravity \cite{Boruch:2023gfn}.\footnote{Such uplifts from $4d$ attractors \cite{Ferrara:1995ih,Ferrara:1996dd} to $5d$ black hole solutions were previously studied in the Lorentzian case in~\cite{Gaiotto:2005gf, Gaiotto:2005xt,Elvang:2005sa,Bena:2005ni}.} 
A diagram of this uplift and all the possible limits that we consider is shown in Figure~\ref{fig:uplift_diagram}. 
The choice of the D6-charge $\Gamma^0$ of the $4d$ black hole determines whether the uplift results in a black hole or a black string. 
If the spatial dimension on which we perform the uplift is the M-theory circle, then by choosing $\Gamma^0 = 1$, we will find a $5d$ black hole saddle in asymptotic $\mathbb R^4 \times S^1$. 
If, instead, we choose $\Gamma^0 = 0$, we find a black string in 
asymptotic $\mathbb R^3 \times S^1_M \times S^1$.\footnote{Choices of $\Gamma^0>1$ result in saddles 
in asymptotic ALE space which we do not discuss in this paper. }

By tuning the temperature of the previously extremal solutions, we can recover previously known extremal solutions. 
In the case of the $5d$ black hole saddles, by taking the size of the thermal circle $\beta_{5d} \to \infty$,  
we recover the $5d$ extremal black hole solutions. 
For $5d$ black string saddles, the limit is more intricate as it also depends on the size $R_M$ of the M-theory circle. 
If we take $\beta_{5d} \to \infty$, while keeping $R_M$ finite, we recover the standard extremal black brane solution 
whose temperature is zero both in the flat space region and in the near-horizon AdS$_3 \times S^2$ region.  
If, instead, we take $\beta_{5d} \to \infty$ while keeping $\beta_{5d}/R_M$ fixed, we recover a black brane saddle 
whose temperature is zero in the asymptotic flat space region but remains finite at the boundary of the 
near-horizon AdS$_3 \times S^2$ region. 
The resulting throat region is described by a complex BTZ black hole saddle that rotates on $S^2$,
which is the true gravitational saddle to the CFT$_2$ elliptic genus, thus filling a gap in the discussion of the 
black hole Farey tail \cite{Dijkgraaf:2000fq}.

The remainder of this paper is organized as follows. In Section~\ref{sec:review}, we review the technical ingredients 
necessary in our calculation. In Section~\ref{section:General_strategy}, we present the detailed relation between 
the $4d$ attractor solutions and the uplifted $5d$ saddles and present the general strategy for the paper. 
We then construct the gravitational saddles for the $5d$ black hole index in Section~\ref{sec:uplifted_5d_black_holes_nonzero_D6} 
and compute their contribution to the index at leading order in $G_N$.  
In Section~\ref{sec:index-5d-black-strings} we construct the saddles for the index of $5d$ black strings, 
analyze the various decoupling limits described above and comment on the relation between the found saddles 
and the elliptic genus of the MSW SCFT$_2$. 
Finally, in Section~\ref{sec:discussion}, we discuss the relation between the newly found saddles and other 
supersymmetric solutions in the literature, we discuss the construction of black rings and other more 
general solutions and discuss how to capture the other sectors for the finite temperature $5d$ gravitational flat space index.

\paragraph{Note added:} During the completion of this work, two interesting articles appeared \cite{Cassani:2024kjn,Adhikari:2024zif}, which also explore the $4d$/$5d$ connection of the attractor saddles and have partial overlap with our results in 
Section~\ref{sec:uplifted_5d_black_holes_nonzero_D6}.

\section{A review of the $4d$ BPS solutions and their gravitational uplift}
\label{sec:review}

\subsection{BPS solutions of $4d$ supergravity \label{subsec:4dsol}}

We begin the discussion with $\mathcal{N}=2$ four-dimensional ungauged supergravity coupled to $\nv$ vector multiplets. 
This theory arises in string theory as the dimensional reduction of 10-dimensional Type IIA supergravity on a Calabi-Yau threefold. 
The bosonic field content of the theory consists of a gravity multiplet ($g_{\mu \nu}, A^0$) involving the $4d$ metric $g_{\mu \nu}$ and 
the graviphoton gauge field\footnote{We will often write gauge fields as one-forms in this paper.} $A^0$, together with $\nv$ vector multiplets $(t^A , A^A)$, $A=1,\dots,\nv$, consisting of complex scalars $t^A=B^A+\ii J^A$ and  gauge fields~$A^A$.

The theory contains $\nv+1$ gauge fields $A^{I} = (A^0, A^A)$, whose electric and magnetic charges form a $(2\nv+2)-$component vector $\Gamma^\alpha = (\Gamma^I , \Gamma_I)$, $\alpha = 1,\dots ,2\nv+2$. 
From the string theory perspective, these vectors are elements of the even cohomology of the Calabi-Yau manifold $H^{2*}(\text{CY}) = H^0 \oplus H^{(1,1)} \oplus H^{(2,2)} \oplus H^{(6)}$ of respective dimensionalities $(1,\nv,\nv,1)$. 
The corresponding charges come from $(D6,D4,D2,D0)$ branes wrapped on appropriate cycles of the CY manifold.
The low-energy theory is invariant under the electric-magnetic duality group~$\text{Sp}(2\nv+2, \mathbb{R})$, which acts as symplectic transformations on the charge vectors~$\Gamma^\alpha$. 
Correspondingly, for two vectors $A,B \in H^{2*}(X) $ there exists a notion of a duality invariant product
\be 
\langle A , B \rangle \= A^I B_I - A_J  B^J \=  
A^\alpha I_{\alpha \beta} B^\beta \, , 
\ee 
which we will refer to as the intersection product. 
We schematically denote the intersection product matrix~$I_{\alpha \beta}$ as
\be 
I_{\alpha \beta} \= 
\begin{pmatrix}
0 &0 & 0 & 1 \\
0 & 0 & 1 & 0 \\
0 & -1 & 0 & 0 \\
-1 & 0 & 0 & 0 
\end{pmatrix}
\,, 
\qquad
I^{\alpha \beta} \= 
\begin{pmatrix}
0 &0 & 0 & -1 \\
0 & 0 & -1 & 0 \\
0 & 1 & 0 & 0 \\
1 & 0 & 0 & 0 
\end{pmatrix}
\, .
\label{eq:intersection_product_matrix}
\ee

It is convenient to parameterize the scalar field space by the so-called projective coordinates~$X^I$, $I=0, \dots, \nv$, 
which are related to the above scalars as~$t^A=X^A/X^0$. 
The two-derivative action is completely determined by the prepotential~$F(X)$, which is a homogeneous function of degree two. 
In this paper, we work with the cubic prepotential of the form 
\be 
F \= \frac{1}{6} D_{ABC} \frac{X^A X^B X^C}{X^0} \,. 
\label{eq:prepotential-in-paper}
\ee
The constants $D_{ABC}$ appearing in the prepotential have the 
interpretation of intersection numbers of the Calabi-Yau compactification manifold. 
The $(2\nv+2)-$component vector~$(X^I , F_I)$,
$F_I \equiv  \partial_I F(X)$ (here and below~$\partial_I \equiv \frac{\partial}{\partial X^I}$),  
also transforms linearly as a vector under the electric-magnetic symplectic transformations. 

The generalized K\"ahler potential~$\CK(X,\overline{X})$ defined as 
\be \label{defK}
e^{-\CK(X,\overline{X})} 
\= \ii \bigl( X^I \overline{F}_I -\overline{X}^I F_I  \bigr) \,,
\ee
is invariant under the duality group and plays an important role in the discussion. 
In particular, it can be used to introduce 
the physical scale
using the above combination as
\be \label{eq:gaugefixingcondition}
e^{-\CK(X,\overline{X})} \= \ell_4^{-2} \,,
\ee
where $\ell_4$ is the $4d$ Planck length. 
(We set~$\ell_4=1$ in the following formulas). This condition gauge fixes the one extra scalar degree of freedom that we introduced above. 
The ``normalized period vector" $\Omega$, given in terms of projective coordinates and the prepotential as 
\be 
\Omega\= e^{\mK/2} (X^I , F_I) \, , 
\qquad 
\langle \Omega, \overline{\Omega} \rangle\= - \ii. 
\ee

We can also express the K\"ahler potential in terms of the scalars~$t^A$ as~$\CK\bigl(X(t),\overline{X}(\overline{t}) \bigr)$ 
evaluated on the hyperplane~\eqref{eq:gaugefixingcondition}. 
This leads to the following expressions,
\be 
\label{eq:Ktrel}
\mathcal{K} \= -\log \left(\frac{4}{3} D_{ABC} J^A J^B J^C \right) \,, \qquad J^A \= \frac{1}{2 \ii} (t^A-\overline{t}^A ) \,,
\ee
from which we find the period vector in terms of $t_A$,
\be 
\Omega \= 
\frac{1}{\sqrt{\frac{4}{3} D_{ABC} J^A J^B J^C}}
 \left(-1 , -t^A , - \frac{t_A^2}{2} , 
\frac{t^3}{6}\right),
\quad
t_A^2 \; \equiv \;  D_{ABC} t^A t^B , \quad 
t^3 \; \equiv  \; D_{ABC} t^A t^B t^C \, . 
\label{eq:normalized_period_vector}
\ee

The bosonic part of the $4d$ action for this theory is given by
\begin{align}
 S &\=  \frac{1}{16\pi} \int \dd^4 x \sqrt{-g} R -  \int 2 G_{A\bar{B}} \, \dd t^A \wedge \ast \dd\Bar{t}^{B} 
\\
& \qquad + \frac{1}{16\pi} \int 
\left( 
\Im \, \CN_{IJ} F^{I} \wedge \ast F^{J} + \Re \, \CN_{IJ} F^{I} \wedge F^{J} 
\right) 
\,.
\label{eq:Lorentzian_action}
\end{align}

The kinetic terms of the gauge fields are functions of the scalars given by the so-called period matrix $(-\CN_{IJ})$, obtained in terms of the second derivatives of the prepotential~$F_{IJ} (X) = \partial_I \partial_J F(X)$ as 
\be
\CN_{IJ}  \= \overline{F}_{IJ} \, + \, \ii \frac{N_{IK} X^K N_{JL} X^L}{X^M N_{MN} X^N} \,, \qquad 
N_{IJ} \=F_{IJ} - \overline{F}_{IJ} \,.
\ee 
The kinetic terms of the scalars are controlled by the K\"ahler metric as  
\be
G_{A \overline B} (t,\overline{t}) \= \frac{\partial^2 \CK}{\partial t^A \, \partial {\overline t}^B} \,,
\ee
with the K\"ahler potential given by the expression~\eqref{eq:Ktrel}.

Now, we turn to BPS solutions. 
The most general supersymmetric black hole solution of $\mathcal{N}=2$ supergravity in four dimensions is described by the metric 
\begin{align}\label{eq:gen4dbh}
\dd s^2_{4d} &\= - \frac{1}{\Sigma(H)} (\dd t + \omega)^2 + \Sigma(H) \dd x^m\dd x^m \, , 
\qquad \bast \, \dd \omega \= \langle \dd H ,H \rangle ,
\\
A^{\alpha} &\= I^{\alpha \beta} \partial_{H^\beta} \log (\Sigma) \left( \dd t+ \omega \right) + \mathcal{A}_d^\alpha \,, 
\qquad 
\dd \mathcal{A}_d^\alpha \= \bast \, \dd H^\alpha \,.
\end{align}
This solution is specified in terms of the
vector $H=(H^0,H^A , H_A , H_0)$ of harmonic functions 
of~$\xvec_a \in \mathbb R^3$ given by 
\be 
H(\xvec) \= \sum_{a} \frac{\gamma_a}{|\xvec - \xvec_a|} + h \,, 
\qquad h \= - 2 \, \Im(e^{-\ii \alpha}\Omega)_{r=\infty} \,.
\ee

The parameter $e^{\ii \alpha}$ is given by
\be 
e^{\ii \alpha}|_{r=\infty} \= \frac{\langle \Gamma_{\text{total}} , \Omega_\infty \rangle}{|\langle \Gamma_{\text{total}}, \Omega_\infty \rangle|} ,
\label{eq:asymptotic_phase_alpha_infty}
\ee
with $\Omega_\infty$ specified in terms of boundary conditions for the scalars in the asymptotic region of flat space. 
For BPS black hole solutions in Lorentzian signature, the residues~$\gamma_a$ of~$H(\xvec)$ simply become the monopole charge~$\Gamma_a$ (i.e., $\gamma_a=\Gamma_a$) 
carried by the black hole at location~$\xvec_a\in \mathbb R^3$. 
Absence of closed timelike curves requires the 
following integrability condition for the one-form field $\omega$,
\be 
 \qquad 
\langle \Gamma_a , H(\xvec_a) \rangle\= 0, \qquad \text{for all poles of }H(\xvec)\text{ located at } \xvec_a \,.
\label{eq:smoothness-Lorentzian}
\ee
Imposing this condition results in the Bates-Denef 
multi-center solutions~\cite{Denef:2000nb,Bates:2003vx,Denef:2007vg}. 
In this paper, we focus, instead, on quasi-Euclidean BPS solutions\footnote{By quasi-Euclidean we mean that the solution could be complex.} for which, as we will see, a weaker condition than~\eqref{eq:smoothness-Lorentzian} must be imposed. 
The solutions that we discuss capture the saddles of the 
gravitational path integral corresponding to the supersymmetric index.
Before we turn to the construction of the Euclidean saddles of the index in Section~\ref{sec:new_attractor_saddles_4d_review}, 
it is useful to analyze the Lorentzian solution a little more, as we do in the rest of this subsection.

The function $\Sigma(H)$ in \eqref{eq:gen4dbh} is in general a homogeneous function of degree~2, 
i.e.~$\Sigma(\lambda H)=\lambda^2 \Sigma(H)$, 
and is commonly referred to as the entropy function, as it controls the size of the horizons. 
For the supersymmetric solutions of interest, it is given explicitly in terms of the period vector 
\be 
\Sigma(H)\= \langle H, \Omega_*(H) \rangle  \langle H , \overline{\Omega}_*(H) \rangle ,
\ee
with $\Omega_*(H)$ defined as the solutions to the generalized attractor equations 
\be 
H(\xvec)\= \ii (\overline{Z}_* (H) \Omega_* (H) - Z_* (H) \overline{\Omega}_* (H)) , 
\ee
which guarantees the existence of Killings spinors in the geometry. 
Note that the above equations imply that $\Omega_*(H)$ is a homogeneous function of degree 0, i.e., $\Omega_*(\lambda H) = \Omega_*(H)$. 
For the cubic 
prepotential \eqref{eq:prepotential-in-paper},  
these equations were solved by Shmakova \cite{Shmakova:1996nz}, and the explicit form of the entropy function is 
\begin{align}
\Sigma(H) &\= \sqrt{Q(H)^3 H^0 - L(H)^2 (H^0)^2} \,, 
\label{eq:entropy_function_Shmakova}
\\ 
L(H)&\= 
-
H_0+ \frac{1}{3} \frac{D_{ABC} H^A H^B H^C}{(H^0)^2} - \frac{H^A H_A}{H^0} , 
\label{eq:L_function_Shmakova}
\\
Q(H)^{3/2} &\= \frac{1}{3} D_{ABC} \, y^A y^B y^C \,, 
\label{eq:Q_function_Shmakova}
\end{align}
where $y^A$ are solutions to the set of algebraic equations
\be 
D_{ABC}\, y(H)^A y(H)^B\= \frac{D_{ABC} H^A H^B}{H^0} - 2 H_C
\label{eq:y_function_Shmakova}
\,.
\ee
Conversely, the full solutions can be explicitly written in terms of the entropy function $\Sigma(H)$ and the 
harmonic functions $H$. In particular, the complex scalars can be written in the form
\begin{align}\label{eq:tfromH}
t^A (H) = 
\frac{\ii H^A + \partial_{H_A}\Sigma}{\ii H^0 + \partial_{H_0}\Sigma}\,.
\end{align}

In this paper, we mostly discuss single-center black holes. The corresponding harmonic function in the Lorentzian theory takes the form 
\be 
H \= \frac{\Gamma}{r} + h \, . 
\ee
Let us briefly describe some features of this solution. 
The extremal entropy of this black hole is simply found by zooming in onto the horizon at $r \rightarrow 0 $,
\be 
S_{\text{ext}} (\Gamma)\= \pi \Sigma(\Gamma). 
\ee
Note that 
the horizon of the extremal black hole is described by a single point $r=0$ in the $\mathbb{R}^3$ base space in the above coordinates.
The pole entering the harmonic function enters the metric as a single second-order pole in the redshift factor $g_{tt}$. 
This corresponds to the fact that near the horizon (as $r \to 0$), the geometry contains an AdS$_2 \times S^2$ throat of infinite proper length. 

We now turn to discuss the Euclidean saddle-point of the index associated with this black hole.

\subsection{The new attractor saddles of the index in $4d$ flat space}
\label{sec:new_attractor_saddles_4d_review}

The general family of Lorentzian supersymmetric solutions to $4d$ supergravity can be used to construct a large family of Euclidean supersymmetric saddles that contribute to the index~\cite{Boruch:2023gfn}, as we now review. 
We only present details in the case of the single black hole relevant to this paper, but the construction can readily be used also for multiple black holes.

\begin{figure}
    \centering
    \includegraphics[width=0.99\linewidth]{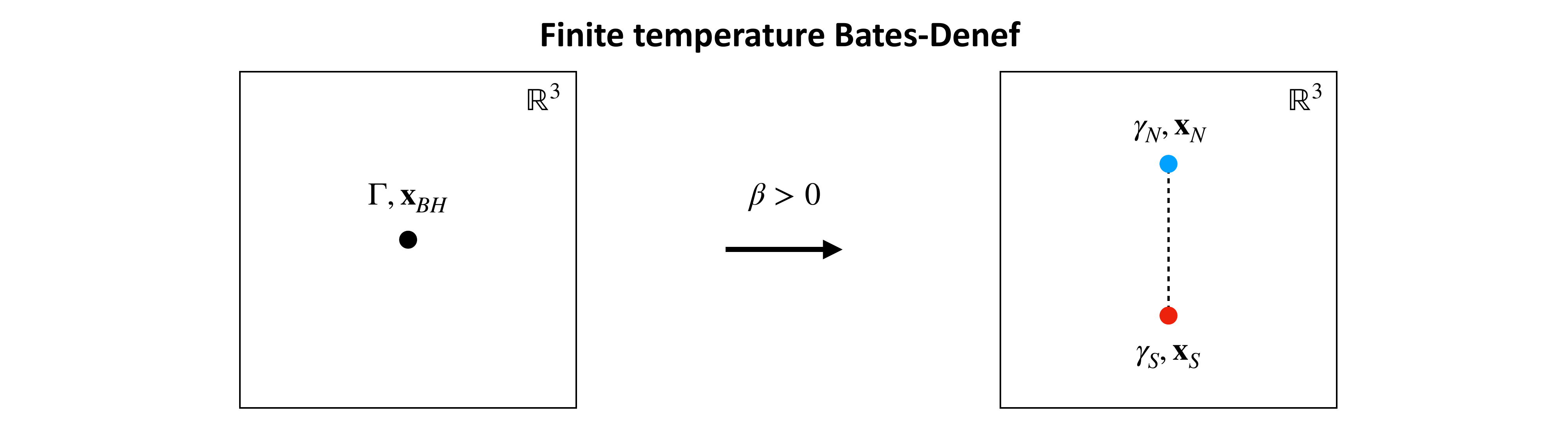}
    \caption{To construct the saddles of the index from extremal supersymmetric solution of total charge $\Gamma$, we split a single double pole $\xvec_{BH}$ of the zero temperature black hole into two simple poles $\xvec_N$ and $\xvec_S$ with respective charges $\gamma_N$, $\gamma_S$, such that $\gamma_N+\gamma_S = \Gamma$. The solution is in the class of Bates-Denef two-center solutions with finite temperature. The condition for the poles to be of first order ensures that the geometry caps off at finite distance and fixes the dipole charge of the black hole, $\gamma_N - \gamma_S$, in terms of total monopole charge $\Gamma$. Requiring smoothness of the new geometry then further fixes the distance between the poles $\xvec_N - \xvec_S$ and at the same time imposes the correct boundary condition for the gravitational supersymmetric index $\beta \Omega_{E,\phi} = 2\pi $. 
    }  \label{fig:finite_temp_Bates_Denef}
\end{figure}

We begin by considering a single extremal black hole of total charge $\Gamma$, described by a harmonic function with a single second-order pole at $\xvec=0$ given at the end of Section~\ref{subsec:4dsol}. 
To turn on a non-zero temperature $T = \beta^{-1}$ (by making the proper length of the time circle to be $\beta$), we go away from extremality by splitting this single second-order pole into two first-order poles located at $\xvec_N$ and $\xvec_S$ (see Figure~\ref{fig:finite_temp_Bates_Denef}), with the total charge kept unchanged. This corresponds to taking
\be 
H \= h + \frac{\gamma_N}{|\xvec- \xvec_N|} + \frac{\gamma_S}{|\xvec- \xvec_S|} ,
\qquad
\gamma_N \; \equiv \; \frac{\Gamma}{2} + \ii \delta , 
\qquad 
\gamma_S \; \equiv \; \frac{\Gamma}{2} - \ii \delta .
\label{eq:4d_attractor_saddle_harmonic_function}
\ee
The condition for the poles at $\xvec_{N/S}$ to be first-order ensures that the geometry caps off at a finite proper distance in the bulk (see~\cite{Boruch:2023gfn} for an extended discussion). Together with the generalized attractor equations, it fixes the dipole charge $\delta$ purely in terms of the total charges $\Gamma$ of the original black hole as
\be
\label{eq:dipole-charge}
\delta \= \frac{1}{2} \left(\barZ_*(\Gamma) \Omega_*(\Gamma) + Z_*(\Gamma) \barOmega_*(\Gamma) \right) \,.
\ee
After redefining the one-form $\omega = -\ii \omega_E$, the Wick-rotated Euclidean metric ($t \rightarrow - \ii t$) takes the form 
\be 
\dd s^2_{4d} \= \frac{1}{\Sigma(H)} (\dd t + \omega_E)^2 + \Sigma(H) \dd x^m\dd x^m \, , 
\qquad 
\bast \, \dd \omega_E \= \ii \langle \dd H , H \rangle \,.
\label{eq:4d_attractor_saddle_metric}
\ee
The gauge fields are given by
\be 
A^{\alpha} \= -\ii I^{\alpha \beta} \partial_{H^\beta} \log (\Sigma) 
\left( \dd t+ \omega_E \right)
+ \mathcal{A}_d^\alpha \,, 
\qquad 
\dd \mathcal{A}_d^\alpha \= \bast \, \dd H^\alpha 
\,,
\label{eq:explicit_gauge_field_A0}
\ee
with the intersection product matrix given in \eqref{eq:intersection_product_matrix}
and the complex scalar fields fixed in terms of harmonic function as in \eqref{eq:tfromH}.
The final step of the construction is to require the smoothness of the geometry on the horizon. 
In our way of writing the metric, this corresponds to the requirement that the Dirac-Misner strings present in the field $\omega_E$ are just coordinate singularities. This turns out to imply 
the simple regularity condition
\begin{align}
\ii \langle \gamma_N , H(\xvec_N)  \rangle\= \frac{\beta}{4\pi} 
\qquad 
\Rightarrow 
\qquad 
|\xvec_N - \xvec_S|\= \frac{\ii \langle \gamma_N,\gamma_S \rangle}{
\frac{\beta}{4\pi} - \langle h, \delta \rangle
}
\,.
\label{eq:north_south_distance}
\end{align}
This condition fixes the base space distance $|\xvec_N - \xvec_S|$ between the north pole and the south pole in terms of the data in the asymptotic boundary conditions ($\beta$ and $\Gamma$).
A useful alternative way of deriving this regularity condition is to introduce ellipsoidal coordinates centered at the poles and require no conical singularities at the tip of the geometry. This imposes the single condition 
\be 
\label{eq:betaOmega}
\beta \, \Omega_{E,\phi}\= 2\pi \,,
\ee 
where $\Omega_{E, \phi}$ is the Euclidean angular velocity of the black hole horizon around the north-south axis of the black hole. 
This is exactly the asymptotic boundary condition required for the grand-canonical function $Z(\beta , \Omega_\phi)$ to compute gravitational supersymmetric index: such an angular velocity inserts
\begin{equation}\label{eq:grandc}
e^{\ii \beta\,\Omega_{E,\Phi} J}=e^{2\pi \ii J} = (-1)^F    
\end{equation}
in the grand-canonical ensemble.  

We can see that these saddles indeed contribute to the $4d$ flat space index by studying the periodicity of fermionic fields around the thermal circle.
When a black hole is rotating, the asymptotic thermal circle is not contractible at the horizon. Instead, the identifications made within the geometry are 
\be
( \tau, \, \phi) \; \sim \; (\tau+\beta, \, \phi- \beta \Omega_{E,\phi}) \; \sim \; (\tau, \phi+2\pi)\,.
\ee
These identifications determine two distinct contractible cycles in the geometry: the first one shrinks to zero at the Euclidean horizon, 
and the second at the axis of rotation. 
Since a fermionic field $\Psi$ has to be anti-periodic when going around a contractible cycle, using \eqref{eq:betaOmega} we find
\be 
\label{eq:fermion-periodicity}
\Psi( \tau, \, \phi)\= - \Psi (\tau+\beta, \, \phi-2\pi)\=-\Psi (\tau, \phi+2\pi) \quad \Rightarrow \quad \Psi(\tau, \,\phi) \= \Psi(\tau+\beta, \phi)\,. 
\ee
This implies that, at the boundary of the spacetime, fermionic fields are indeed periodic around the thermal circle (which is not contractible in the bulk).  
This is
precisely what is expected of a supersymmetric index.

\subsection{The $4d$/$5d$ lift}
\label{sec:4d/5d-uplift-review}

The $4d$/$5d$ lift \cite{Behrndt:2005he,Gaiotto:2005gf,Gaiotto:2005xt,Elvang:2005sa,Bena:2005ni, Banerjee:2011ts} relates solutions of the $4d$ theory obtained from compactifying type IIA supergravity on a CY manifold \cite{Bodner:1990zm}
with solutions of the $5d$ theory obtained from 11d supergravity on the same CY manifold \cite{Cadavid:1995bk,Kraus:2005gh,Larsen:2006xm}. 
The bosonic part of the relevant $5d$ action has the simple form
\begin{align}
16\pi \ell_5^3 \,I_5\= 
 &\int \left( R_{(5)} \ast_5 1
+ \frac{1}{2} D_{ABC} Y^C
\dd Y^A \wedge 
\ast_5 \dd Y^B 
+ a_{AB} F^A_{5d} \wedge \ast_5 F^B_{5d}
\right)
\\
& \nonumber+  \int \frac{\sqrt{2}}{3} D_{ABC} F^A_{5d} \wedge F^B_{5d} \wedge A^C_{5d} 
,
\end{align}
where
\be 
a_{AB}\= 
D_{ABC} Y^C 
- \frac{1}{4} D_{ACD} Y^C Y^D 
D_{BEF} Y^E Y^F ,
\ee 
and the $5d$ scalars are subject to the condition 
\be 
D_{ABC} Y^A Y^B Y^C = 6 .
\ee
This action reduces to the 4$d$ action \eqref{eq:Lorentzian_action} after inserting the following ansatz, known as the uplift formula,\footnote{For details of the reduction, see e.g. \cite{Looyestijn:2010pb}.}
\begin{align}
\frac{\dd s_{5d}^2}{\ell_5^2} &\= (2\widetilde{V}_\IIA)^{2/3} \, (\dd\psi+A^0)^2 + (2\widetilde{V}_\IIA)^{-1/3} 
\, {\dd s^2_{4d}} \,, \label{eq:5d_uplift_formula_1}
\\  
{A^A_{5d}}&\= -A^A + \Re t^A \, (\dd\psi + A^0) \,, 
\qquad 
Y^A\= \frac{\Im t^A}{\widetilde{V}_\IIA^{1/3}}
\,, \label{eq:5d_uplift_formula_2}
\end{align}
where $\dd s^2_{4d}$ is the Einstein-frame $4d$ metric \eqref{eq:4d_attractor_saddle_metric} described above in the convention where $\ell_4=1$. Restoring $\ell_4$, with the previous ansatz the $5d$ Planck length $\ell_5$ is set to be 
\be 
\frac{\ell_5^2}{\ell_4^2} = \frac{R_M}{\ell_5} = 4\pi  (2 \widetilde V_{\IIA, \infty})^{1/3}  \,.
\label{eq:ell-5-over-ell-4}
\ee
When using the $4d$ Lorentzian solution described in section \ref{subsec:4dsol}, the resulting $5d$ solution is an extremal $5d$ black hole. 
We will instead be interested in uplifting the new attractor solutions described in Section~\ref{sec:new_attractor_saddles_4d_review} to compute the index of $5d$ black holes and black strings directly from the gravitational path integral. 
Before we explain the general strategy for this uplift, let us describe some of our $5d$ conventions. 
In \eqref{eq:5d_uplift_formula_1} and \eqref{eq:5d_uplift_formula_2}
$\widetilde{V}_{\text{IIA}} $ denotes the space-dependent dimensionless Calabi-Yau volume measured in string units
\be 
\widetilde{V}_{\text{IIA}} 
\equiv \frac{1}{6} D_{ABC} J^A J^B J^C
\= \frac{1}{2} \left(
\frac{\Sigma}{H^0 Q}
\right)^3 .
\label{eq:CY_volume_space_dependent}
\ee
The right-hand side of the equation follows from
\be 
\Im t^A \= J^A\= 
\frac{y^A \Sigma}{H^0 Q^{3/2}} \,, 
\ee
which can be derived from the explicit solution for the scalars $t^A (H)$ written above.
The relation~\eqref{eq:CY_volume_space_dependent} 
leads to many simplifications in the metric. 
When~$H^0 \neq 0$, it becomes natural to rewrite the metric in terms of the functions $Q$ and $L$ entering the entropy function 
(see eqs.~\eqref{eq:entropy_function_Shmakova}--\eqref{eq:y_function_Shmakova}) as 
\begin{align}
\label{eq:5dmetricQL}
\frac{\dd s^2_{5d}}{\ell_5^2} &\= 
- \frac{1}{Q^2} \left(
\dd t + \omega 
+ L(\dd \psi + \mathcal{A}_d^0) \right)^2 
+Q \, \dd s^2_{\text{TN}}  \,,  
\\
{\dd s^2_{\text{TN}}}
&\=
\frac{1}{H^0} 
\left(\dd \psi + \mathcal{A}_d^0\right)^2 
+ H^0 \, \dd x^m \dd x^m 
\, .
\end{align}
This allows us to interpret the resulting metric in terms of black holes/black rings sitting on a Taub-NUT background. 
The $4d$/$5d$ lift applies only to a restricted family of $5d$ solutions since all the uplifted saddles contain at least the $U(1)$ isometry corresponding to the $\psi$-circle.

\section{The $5d$ solutions and their on-shell action }
\label{section:General_strategy}

In this section, we describe the strategy for constructing the black hole and black string saddles of the $5d$ gravitational index. 
We start with Euclidean gravitational saddles with asymptotic metric $S^1_{\beta_{5d} }\times S^1_{M} \times \mathbb R^3$ and corresponding 
asymptotic symmetry $U(1)_{t_E} \times U(1)_L \times SU(2)_R$.
The solutions are specified by the following parameters:
\begin{itemize}
    \item Fixed proper length of the thermal circle $S^1_{\beta}$ to $\beta_{5d} = \sqrt{4\pi} \beta$; more specifically, we shall fix $\beta_{5d}/\ell_5$.\footnote{The relation $\beta_{5d} = \sqrt{4\pi} \beta$ comes from computing the proper lengths of the $5d$ thermal circle $\beta_{5d}$ and $4d$ thermal circle $\beta$ using \eqref{eq:5d_uplift_formula_1} and the ratio $\ell_5/\ell_4$ \eqref{eq:ell-5-over-ell-4}. }
    \item Fixed size $R_M/\ell_5$ for the M-theory circle $S^1_{M}$. We will also consider the limit $R_M/\ell_5 \to \infty$ in which the M-theory circle is decompactified. 
    In this case the saddles have the asymptotic metric  $S^1_{\beta_{5d}} \times \mathbb R^4$, and symmetry gets enhanced to~$U(1)_{t_E} \times SU(2)_L \times SU(2)_R$.
    \item Fixed values for the scalar moduli $Y^A$. 
    \item  Fixed asymptotic electric and magnetic charges (given by the vector $\Gamma_{5d}$), and components of all gauge fields along the $\psi$ direction.\footnote{We normalize $\Gamma_{5d}$ such that all components of the vector are integer-valued and independent of the $5d$ electromagnetic coupling constant. }
    \item Fixed~$J_L$. Here $J_L$ is the momentum around the M-theory circle or, equivalently, the charge under the~$U(1)_L$ symmetry. When the symmetry is enhanced to~$SU(2)_L$, $J_L$ is the angular momentum in the Cartan subalgebra of the~$SU(2)_L$. 
    \item Fixed angular velocity $\Omega_R$ associated with the Cartan generator of~$SU(2)_R$.\footnote{The corresponding angular momentum~$J_R$ is half-integer quantized.} The Euclidean  angular velocity is fixed to $\Omega_R \, \beta = 2\pi $, which implements the insertion of~$e^{2\pi \ii J} = (-1)^F$ in the trace. These are the uplifts of conditions \eqref{eq:betaOmega} and \eqref{eq:grandc}. Equivalently, we impose that all fermionic fields are periodic around the thermal circle. 
\end{itemize}

To construct these solutions, we start with the Euclidean version of the $5d$ uplift formula \eqref{eq:5d_uplift_formula_1},
\be
\label{eq:Euclidean-metric-uplift}
\frac{\dd s_{5d}^2}{\ell_5^2}\= (2\widetilde{V}_\IIA)^{2/3} \, (\dd\psi+A^0)^2 + (2\widetilde{V}_\IIA)^{-1/3} 
\, \dd s^2_{4d} \,, \quad \text{ with }\quad  \psi \;\sim \; \psi+4\pi\,,
\ee
where $\psi$ parametrizes the M-theory circle $S^1_M$, while the thermal circle $S^1_{\beta_{5d} }$ is part of $\dd s^2_{4d}$. 

Recall that the action of $5d$ supergravity reduces to the $4d$ supergravity action with one KK vector multiplet. 
For the ansatz~\eqref{eq:Euclidean-metric-uplift}, the scalar in the KK vector multiplet is responsible for setting the size of the M-theory circle.
The electric component of the KK gauge field determines the charge~$J_L$ of the solution, while the magnetic component determines the fibration of the $S^1_M$ around the $4d$ base space.  
Using the new attractor BPS saddles discussed in Section~\ref{sec:new_attractor_saddles_4d_review} for the $4d$ metric $\dd s^2_{4d}$, 
we obtain the class of smooth $5d$ BPS solutions that have~$U(1)_{t} \times U(1)_L \times U(1)_R$ isometry. 
Asymptotically, the $(-1)^F$ is a central element of $SU(2)_R$, which is why  the asymptotic symmetry of our solutions is $U(1)_{t} \times U(1)_L \times SU(2)_R$. 
The dipole charge that is turned on in the solution spontaneously breaks the~$SU(2)_R$ asymptotic symmetry down to~$U(1)_R$. 
As a result of this breaking, there are two Goldstone modes that rotate the axis of the dipole but leave the asymptotic boundary conditions unchanged~\cite{H:2023qko}.

The boundary conditions for the $5d$ gravitational index saddles determine the boundary conditions in~$4d$ in our construction. 
The length of the M-theory circle fixes the value of $\widetilde V_{\IIA}$ at the asymptotic boundary ($\widetilde V_{\IIA,\infty}$) according to \eqref{eq:ell-5-over-ell-4}.  
Similarly, the values of the $5d$ scalars $Y^A$ determine the value of the $4d$ scalars $t^A$ at the boundary. 
The components of the charge vector $\Gamma_{5d}$ can be obtained from the components of $\Gamma$~\cite{Gaiotto:2005gf}. 
Since $\Gamma_{5d}$ has $2\nv$ components, while $\Gamma$ has $2\nv+2$ components, there are two remaining components of $\Gamma$ that need to be fixed. These are the components that acquire a geometric/gravitational interpretation in $5d$. 
The first component~$\Gamma_0$ corresponds to angular momentum $J_L$. The other component~$\Gamma^0$ is, as we shall see shortly,  a nut charge that determines the topology of the Euclidean horizon of the $5d$ saddle, 
dictating whether we see a $5d$ black hole or a $5d$ black string. 
Since the relation of the $5d$ charges to the $4d$ charges $\Gamma$ depends on the choice of  $\Gamma^0$, correspondingly, 
we give separate presentations for black holes and black strings.

Finally, the Euclidean $5d$ angular velocity $\Omega_R$ is fixed to $\Omega_R \, \beta_{5d} = 2\pi$ which follows from fixing the $4d$ angular velocity in the index attractor saddles as in \eqref{eq:betaOmega}. 
Relatedly, since fermionic fields are periodic around the thermal circle in $4d$ as seen in \eqref{eq:fermion-periodicity}, they are also periodic under translations around the thermal circle in~$5d$ combined with an appropriate shift on $S^1_M$. 
Consequently, the $5d$ solution~\eqref{eq:Euclidean-metric-uplift} is a saddle for the~$5d$ gravitational supersymmetric index.

What is the contribution of this saddle to the index? 
Recall from Section~\ref{sec:4d/5d-uplift-review} that the $5d$ action reduces to the $4d$ action under the above reduction on~$S^1_M$, and hence the semi-classical contribution of our $5d$ saddles can be found simply
by evaluating the Euclidean on-shell action of the $4d$ attractor saddles. 
Using that the throat is finite and that there are no singularities at the poles of the dipole, the answer for the $4d$ on-shell action \cite{Boruch:2023gfn} is simple,
\be 
-\widetilde I^{\text{on-shell}}_E(\beta, \Omega, \Gamma)
\= -\beta |Z(\Gamma ; \Omega_\infty)|  +  
\pi 
\Sigma(\Gamma)
\=-\beta \, M_{BPS}(\Gamma ; \Omega_\infty)  + S_{BH}^{\text{extremal}}
\, .
 \label{eq:index-invariant}
\ee

Even though this is a finite-temperature black hole saddle, 
the regularity of the geometry leads to nontrivial cancellations, which fixes the final contribution of the saddle to simply be given by a sum of two terms: 
(i) a Boltzmann term with trivial temperature dependence and with an energy that saturates the BPS bound, and 
(ii) the zero-temperature entropy of the corresponding extremal black hole. 
Since we fix the charges of the black hole and the values of the moduli through our boundary conditions, we can add a charge-dependent 
boundary counterterm to the action, with vanishing variation, such that it subtracts the Boltzmann term.  
After adding this counterterm, the on-shell action is 
\be 
-I^{\text{on-shell}}_E(\beta, \Omega, \Gamma)
\=  
\pi 
\Sigma(\Gamma) =S_{BH}^{\text{extremal}}=\log(d_b - d_f)  
\, .
 \label{eq:index-invariant2}
\ee
Using the relation~\eqref{eq:index-invariant2} we can write  $I^{\text{on-shell}}_E$ in terms of the asymptotic charges $\Gamma_{5d}$, the angular momentum component $J_R$ and the charge $\Gamma^0$. 
Since \eqref{eq:index-invariant2} is now completely independent of either the $5d$ or $4d$ temperatures and moduli, we find a continuous family of saddles, all of which have the same on-shell action; this agrees with our expectation that the boundary index is independent of the temperature and moduli and is equal (at least at leading order) in $4d$ and $5d$.
Only in the zero temperature limit do the saddles in these families become  
the extremal solutions whose entropy is the same as the index in \eqref{eq:index-invariant2}.

Depending on the choice of the charge~$\Gamma^0$ we can distinguish two kinds of saddles, whose differences are summarized in Figure~\ref{fig:uplift_diagram}:
\begin{itemize}

    \item When $\Gamma^0>1$ we obtain a $5d$ asymptotically ALE space that takes the form $\mathbb R^4/\mathbb Z_{\Gamma^0} \times S^1_\beta$. While such solutions are interesting in their own right, we will not focus on them since they do not contribute to the index of black holes or black strings, at least when considering single-center contributions. 
  
    \item \textbf{$5d$ black hole.} When~$\Gamma^0=1$, the  uplift of the $4d$ attractor saddle leads to a~$5d$ black hole solution. Its horizon topology is $S^3$ due to the non-trivial fibration of $S^1$ over the $S^2$ horizon of the $4d$ attractor saddle. 
    In this case, we can also consider a decompactification limit, in which we take the size of the M-theory circle $R_M/\ell_5 \to \infty$, keeping the $5d$ temperature $\beta/\ell_5$ finite. 
    Using the fact that the on-shell action is independent of the $4d$ moduli and $4d$ temperature, we thus find solutions that contribute to the index with $\mathbb R^4 \times S^1_\beta$ asymptotics, whose contribution is independent of the $5d$ temperature and $5d$ moduli.    
  
    \item \textbf{$5d$ black string.} When~$\Gamma^0 = 0$, the uplift of the $4d$ black hole solution leads to a black string whose horizon topology is $S^2 \times S^1$. 
    Although the simple uplift may look trivial, there are two interesting new regimes for these solutions that we study. 
    The first is the finite temperature regime in which we keep $\beta/\ell_5$ and $R_M/\ell_5$ finite. 
    In this case, the black string solution does \textit{not} have an infinitely long throat and is not decoupled from the asymptotic boundary.     
    In the second regime, we take the inverse temperature of the~$5d$ black string $\beta/\ell_5$ to infinity 
    while appropriately scaling the moduli in such a way that $\beta/R_M$ remains finite. 
    In this regime, the near-horizon region of the black string is a rotating BTZ black hole that captures the gravitational index in asymptotically AdS$_3 \times S^2$. 
    Here, the gravitational index corresponds to the elliptic genus of the dual MSW CFT$_2$. 
\end{itemize}

It is worth noting that the black string leads to two saddles that correspond to different observables in different theories.
On the one hand, we have the decoupled (complex) finite-temperature spinning BTZ~$\times S^2$ saddle of the SCFT$_2$ elliptic genus,
and on the other hand we have the finite-temperature supersymmetric black string saddle of the corresponding index in $5d$ flat space---with no decoupling limit.
From the point of view of the microscopic index they are related but not the same: the elliptic 
genus of the black string is one sector of the index of $5d$ flat space,\footnote{This is the sector captured by a single SCFT$_2$, whose gravity dual is the single-center black brane saddle. } which, more generally, takes into account 
all possible supersymmetric objects and their excitations in asymptotically flat $5d$.
What is remarkable is that we have found the precise gravitational dual of this microscopic intuition. 
Indeed, the family of solutions we have found for the gravitational index allows us to 
interpolate between these two ends without changing the leading value of the saddle.

\section{The gravitational index of $5d$ black holes}
\label{sec:uplifted_5d_black_holes_nonzero_D6}

In this section, we analyze the explicit form of the uplifted $5d$ index saddles for the case of nonzero total D6 charge $\Gamma^0 \neq 0$. 
We begin by discussing the scaling of the scalar moduli, which allows for the decompactification of the M-theory circle. 
The resulting solution is a saddle of five-dimensional supergravity in asymptotically flat space. 
We verify that it is smooth and that it satisfies appropriate boundary conditions at infinity 
corresponding to the~$5d$ gravitational index.
Further, we verify in a simple example that the new family of saddles generalizes 
the saddles of the $5d$ index recently constructed in~\cite{Anupam:2023yns} to the case of an arbitrary number of vector multiplets. 

\subsection{Decompactification limit and the $5d$ index saddle}

Our starting point is the $4d$ attractor saddle at a generic point in the scalar moduli space 
\be
t^A|_\infty \; \equiv \; b^A + \ii j^A\,, 
\ee
with total asymptotic charge 
\be 
\Gamma \= (\Gamma^0 > 0 \,, \, \Gamma^A \, , \, \Gamma_A ,\, \, \Gamma_0) 
\, .
\ee
As explained in Section~\ref{sec:new_attractor_saddles_4d_review}, this saddle is described by the harmonic function~\eqref{eq:4d_attractor_saddle_harmonic_function} with north pole~$(\xvec_N,\gamma_N=\Gamma/2+\ii \delta(\Gamma))$ and south pole~$(\xvec_S,\gamma_S =\Gamma/2-\ii \delta(\Gamma))$. 
The $\beta$-dependent distance $|\xvec_N - \xvec_S|$ is given by~\eqref{eq:north_south_distance}, and the explicit formula for the dipole charges $\delta(\Gamma)$ in terms of the charge~$\Gamma$ are given  in~\eqref{eq:explicit_dipole_charge_eq1} and \eqref{eq:explicit_dipole_charge_eq2}. 
After uplifting this solution, the resulting $5d$ Euclidean metric is now simply given in terms of \eqref{eq:5d_uplift_formula_1} with the $4d$ metric given by \eqref{eq:4d_attractor_saddle_metric} and the CY volume given by \eqref{eq:CY_volume_space_dependent}. 
The components of the constant $h$ in the harmonic functions can be determined in terms of asymptotic charge and scalar moduli from~\eqref{eq:normalized_period_vector}, \eqref{eq:asymptotic_phase_alpha_infty} 
(explicit expressions can also be found in \cite{Cheng:2008gx}). 
Recalling that, once we specify the entropy function, the metric is determined in terms of the harmonic functions $H(\xvec)$, 
we see that the above data fully specifies a Euclidean solution to $5d$ supergravity with $S^1_\beta \times S^1_M \times \mathbb{R}^3$ asymptotics. 

To obtain a saddle in $5d$ asymptotically flat space at finite temperature, we need to perform the further step of decompactifying the M-theory circle, i.e.~take $R_M/\ell_5 \rightarrow \infty$. 
Here, it is important to remember that the proper length of the M-theory circle is related to the scalar moduli at asymptotic infinity as in \eqref{eq:ell-5-over-ell-4}, which can be rewritten as 
\be 
\frac{R_M}{4\pi} 
\= \ell_5 \left(\frac{1}{3}D_{ABC}j^A j^B j^C \right)^{1/3} 
\,.
\label{eq:relation_R_M_and_moduli}
\ee
Therefore, one way of obtaining the limit to~$5d$ flat space is to first set
the following boundary conditions, 
\be 
b^A \; \sim \; O(1) \,, \qquad j^A \;\sim \; \Lambda^2  \qquad 
\Longrightarrow \qquad R_M/\ell_5 \;\sim \; \Lambda^2 \,,
\qquad \ell_5/\ell_4 \; \sim \; \Lambda \,. 
\label{eq:parameters_scaling_Lambda}
\ee
Taking the limit $\Lambda \rightarrow \infty$, 
then makes the size of the M-theory circle measured in $5d$ Planck units infinite, whilst keeping finite the inverse temperature measured in $5d$ Planck units, $\beta/\ell_5$. 

The parameter $\Lambda$ now enters the metric through the constant~$h$, whose leading behavior for~$\Lambda \gg 1$ is
\be 
h \=  \frac{1}{\sqrt{\frac{4}{3} D_{ABC} j^A j^B j^C  }} 
\left( 0 \,,\, 0\,,\, -  D_{ABC} j^B j^C  
\,,\, \frac{1}{\Gamma^0} D_{ABC} \Gamma^A j^B j^C  
\right) 
\; \sim \; ( 0,0, \Lambda , \Lambda )
\,.
\ee
This implies that as we take the decompactification limit, the interesting part of the geometry localizes to coordinate regions of order $1/\Lambda$. 
To zoom in on this part of the metric, we rescale the spatial coordinates as~$\xvec \rightarrow \xvec/\Lambda$, 
in terms of which the harmonic function now scales uniformly as
\be 
H(h,\xvec) \= h + \frac{\gamma_N}{r_N} + \frac{\gamma_S}{r_S} 
\= \Lambda 
\left( \frac{h}{\Lambda} + 
\frac{\gamma_N}{\Lambda r_N} + 
\frac{\gamma_S}{\Lambda r_S}
\right)
\= \Lambda \, H(h/\Lambda, \xvec/\Lambda) \,. 
\ee

Because the entropy function is a homogeneous function of order 2, i.e.~$\Sigma(\Lambda H) = \Lambda^2 \Sigma(H)$, 
it is natural to redefine the time coordinate 
in~\eqref{eq:4d_attractor_saddle_metric}
as~$t \rightarrow \Lambda t $. 
This change of coordinates is consistent with the scaling of the proper length of the asymptotic time circle as
\be 
\label{eq:definition-tilde-beta-1}
 \widetilde \beta \; \equiv \; \frac{\beta} {  \Lambda 
 \ell_4}  \= \frac{\beta_{5d}}{\ell_5 }
 \frac{(2\widetilde V_{\IIA, \infty})^{1/6}}{\Lambda}
 \,,
\ee 
where we have introduced the dimensionless parameter $\widetilde \beta$, which remains fixed and finite in the limit. 
As we take $\Lambda \to \infty$, 
the $4d$ saddle has zero temperature measured in units of $\ell_4$ while its $5d$ uplift has a non-zero temperature measured in units of $\ell_5$.

The formulas below are simpler to write in terms of the parameter~$\widetilde \beta$. Using \eqref{eq:north_south_distance}, 
one can easily check that the distance between the north pole and the south pole in the base space, in rescaled coordinates, is
\be 
|\xvec_N - \xvec_S| \= \frac{\ii \langle \gamma_N,\gamma_S \rangle}{
\frac{\widetilde \beta}{4\pi} - \langle h, \delta \rangle
} \,.
\ee
This stays invariant as we take the decompatification limit $\Lambda \rightarrow \infty$.

Having implemented the boundary conditions and coordinate changes as above, we can now take the strict limit $\Lambda \rightarrow \infty$ and obtain the metric of the $5d$ Euclidean saddle at finite temperature.
This saddle has a Euclidean time circle $t \sim t+\widetilde \beta$ and four non-compact spatial directions. 
The metric can again be written in terms of the original uplift formula, \eqref{eq:5d_uplift_formula_1}, with the important difference that now 
the only nonzero components of~$h$ are  $h_A, h_0$
\be 
h \= \frac{1}{\Lambda \sqrt{\frac{4}{3}j^3}} 
\left( 0 , 0, - \frac{\Gamma^0}{\Gamma^0} j_A^2 
, \frac{1}{\Gamma^0} \Gamma^A j_A^2  
\right) 
\= \left( 
0,0, h_A , - \frac{\Gamma^A}{\Gamma^0} 
h_A
\right)\,
\ee
(recall that $j^A \sim \Lambda^2$). The branes that do not survive the decompatification limit are those that wrap the M-theory circle, namely the KK monopoles and the M5 branes.

Before analyzing the smoothness and fixing the asymptotics of the new saddles, let us write down the $5d$ solutions explicitly. 
The Euclidean form of the uplifted metric~\eqref{eq:5dmetricQL} is
\begin{align}
\label{eq:5d_uplifted_BH_metric_Start}
\dd s^2_{5d} &\= 
\frac{1}{Q^2} \left(
\dd t + \omega_E 
+ \ii L(\dd \psi + \mathcal{A}_d^0) \right)^2 
+Q \dd s^2_{\text{TN}} \,,  
&
\bast \, \dd \omega_E &\= \ii \langle \dd H ,H \rangle \,,
\\
\dd s^2_{\text{TN}}
&\=
\frac{1}{H^0} 
(\dd \psi + \mathcal{A}_d^0)^2 
+ H^0 \dd x^m \dd x^m 
\,, 
& \bast \, \dd \mathcal{A}_d^0 
&\= \dd H^0 \,.
\end{align}
Here, the functions $Q$ and $L$ are given by~\eqref{eq:Q_function_Shmakova}, \eqref{eq:L_function_Shmakova}, \eqref{eq:y_function_Shmakova}, 
in terms of the harmonic function (with~$r_i \equiv |\xvec - \xvec_i|$)
\begin{align}
H^0 &\= \frac{\Gamma^0}{2} \left( 
\frac{1}{r_N} + \frac{1}{r_S}
\right)   
+\ii \delta^0 \left( 
\frac{1}{r_N} - \frac{1}{r_S}
\right)
,
\\ 
H^A &\= \frac{\Gamma^A}{2} \left( 
\frac{1}{r_N} + \frac{1}{r_S}
\right)   
+\ii \delta^A \left( 
\frac{1}{r_N} - \frac{1}{r_S}
\right)
,
\\ 
H_A &\= h_A + \frac{\Gamma_A}{2} \left( 
\frac{1}{r_N} + \frac{1}{r_S}
\right)   
+\ii \delta_A \left( 
\frac{1}{r_N} - \frac{1}{r_S}
\right)
,
\\ 
H_0 &\= h_0 + \frac{\Gamma_0}{2} \left( 
\frac{1}{r_N} + \frac{1}{r_S}
\right)   
+\ii \delta_0 \left( 
\frac{1}{r_N} - \frac{1}{r_S}
\right)
.
\end{align}
The equations for the one-form fields can be explicitly solved and yield
\begin{align}
\omega_E &\= 
\langle h , \delta \rangle
(\hat{\omega}_N - \hat{\omega}_{S} ) 
+ \ii \langle
\gamma_N , \gamma_S
\rangle \hat{\omega}_{NS} \, , 
\\
\mathcal{A}_d^0 &\= 
\gamma^0_N \hat{\omega}_N 
+ \gamma^0_S \hat{\omega}_S \=
\frac{\Gamma^0}{2} (\hat{\omega}_N + \hat{\omega}_S)
+
\ii \delta^0 (\hat{\omega}_N - \hat{\omega}_S)\, ,
\label{eq:explicit_gauge_field_Ad0}
\end{align}
in terms of functions $\hat{\omega}_N,\hat{\omega}_S,\hat{\omega}_{NS}$, defined as solutions to~\cite{Katona:2023uaj}
\begin{align}
\bast \, \dd \hat{\omega}_N 
&\= \dd \left( 
\frac{1}{r_N}
\right)  , 
\qquad
\bast \, \dd \hat{\omega}_S 
\= \dd \left( 
\frac{1}{r_S}
\right) ,
\\ 
\bast \, \dd \hat{\omega}_{NS}
&\= 
\frac{1}{r_S} \dd 
\left( 
\frac{1}{r_N}
\right) 
-
\frac{1}{r_N} \dd 
\left( 
\frac{1}{r_S}
\right)  .
\label{eq:5d_uplifted_BH_metric_End}
\end{align}

\paragraph{Asymptotics:} We now impose that the metric is asymptotically $S^1_{\widetilde \beta} \times \mathbb{R}^4$. To this end, we expand the harmonic functions at asymptotic infinity, which leads to the following behaviors of functions entering the metric 
\begin{align}
Q^{3/2}_\infty &\= \frac{1}{3} D_{ABC} y_\infty^A y_\infty^B y_\infty^C 
+ O(r^{-1})\,, 
\qquad 
D_{ABC} y_\infty^A y_\infty^B \= - 2 h_C \,,
\\ 
\omega_E &\= 
\frac{2\ii J_R \sin^2 \theta \dd \phi}{r} 
+ 
O(r^{-3})
\,,
\\ 
L &\= \frac{1}{r}
\left( 
\frac{D_{ABC}\Gamma^A \Gamma^B \Gamma^C}{3 (\Gamma^0)^2} 
- \frac{\Gamma^A \Gamma_A}{\Gamma^0}
- \Gamma_0
\right) 
+ \frac{2}{\Gamma^0} \frac{J_R \cos \theta}{r} 
+ O\left(r^{-2} \right)
, 
\\ 
&\=  \frac{2 J_L}{r} 
+ \frac{2}{\Gamma^0} \frac{J_R \cos \theta}{r} 
+ O\left(r^{-2} \right) ,
\\
\mathcal{A}_d^0 &\= \Gamma^0 \cos \theta \dd \phi +  O\left( r^{-1} \right) .
\end{align}
From this, we derive that, to leading order, the metric behaves as
\be 
\dd s^2 |_\infty \= \dd t'^2 + \dd \rho^2 + 
\frac{\rho^2}{4} \left( 
\frac{(\dd \psi  +\Gamma^0 \cos \theta \dd \phi)^2}{(\Gamma^0)^2} + \dd \Omega_2^2 
\right) + \dots , 
\ee
where we have rescaled the time coordinate and redefined the radial coordinate as
\be 
t \= Q_\infty t' , 
\qquad 
r \= \frac{\rho^2}{4} \frac{1}{\Gamma^0 Q_\infty} .
\ee
Thus, as mentioned in Section~\ref{section:General_strategy},  we see that asymptotically the metric is $\mathbb{R}^4/\mathbb{Z}_{\Gamma^0} \times S^1_{\widetilde \beta}$
and if we want to work with an asymptotically globally flat metric, we must impose $\Gamma^0 = 1$.

\paragraph{Topology and smoothness:}
Let us start by describing the topology of the horizon. As we have shown above, the induced metric on the fixed $\rho$ surface is asymptotically  $S^3$. 
It is useful to introduce  {ellipsoidal} coordinates
\begin{align}
x &\= \frac{r_{NS}}{2} \sinh{\mu} \sin \theta \cos \phi\,, 
\,\,\,\,\,\,
y \=  - \frac{r_{NS}}{2} \sinh{\mu} \sin \theta \sin \phi
\,, 
\,\,\,\,\,\,
z \= \frac{r_{NS}}{2}\cosh{\mu} \cos \theta\,, 
\\ 
\xvec_N &\= \left(0,0, \frac{r_{NS}}{2} \right) , 
\qquad 
\xvec_S \=  \left(0,0, -\frac{r_{NS}}{2} \right),
\end{align}
in which the horizon is the surface~$\mu=0$. 
By increasing $\mu>0$, we can move the surface surrounding the horizon to infinity without crossing any poles of the metric. 
In this way, we can continuously deform the surface infinitesimally close to the horizon to the asymptotic~$S^3$ described above, thus showing that the horizon is topologically~$S^3$.

The $S^3$ horizon of the uplifted black hole rotates now in the $\psi$-direction with an angular velocity determined by the D6 dipole charge $\delta^0$. This can be seen by introducing corotating coordinates 
\be 
\psi_{\text{cor}} \= \psi - \ii \Omega_\psi t \,, 
\qquad 
\phi_{\text{cor}} \= \phi - \ii \Omega_\phi t \,, 
\ee
and requiring that the induced metric on the horizon is static. 
Using the explicit expression for the gauge fields \eqref{eq:explicit_gauge_field_A0}, \eqref{eq:explicit_gauge_field_Ad0}, and the fact that $\omega_{E,\phi}|_{\text{hor}} = \frac{\ii}{\Omega_\phi} = \frac{\widetilde \beta}{2\pi}$ one finds that 
\be 
\dd \psi + A^0|_{\text{horizon}} 
\= 
\dd \psi_{\text{cor}} + \ii \dd t (\Omega_\psi 
+ a_\phi \Omega_\phi ) + \dots ,
\ee
where we defined $\mathcal{A}_d^0|_{\text{horizon}} \equiv a_\phi \dd \phi = a_\phi \dd \phi_{\text{cor}} + a_\phi \ii \Omega_\phi \dd t$. Furthermore, on the horizon we find that
\be 
\mathcal{A}_d^0|_{\text{horizon}} 
\; \equiv \; a_\phi \dd \phi
\= 
\frac{1}{2}
(\hat{\omega}_N + \hat{\omega}_S) + \ii \delta^0 (\hat{\omega}_N - \hat{\omega}_S) 
|_{\text{horizon}} 
\= 2\ii \delta^0 \dd \phi  
\,,
\ee
from which we find that the classical value of angular velocity in $\psi$-direction is proportional to D6 dipole charge 
\be 
\Omega_\psi \= - a_\phi \Omega_\phi \= \frac{4\pi \delta^0}{\widetilde \beta} \,.
\ee

To analyze the smoothness, it suffices to focus on the region near the poles, since the metric in the rest of the space is manifestly smooth. 
For this purpose, it turns out to be simpler to work with the original uplift formula~\eqref{eq:5d_uplift_formula_1}. 
Close to the north pole (the analysis for the south pole is identical), we have that
\begin{align}
\omega|_{r_N} & \; \simeq \; -
\left(
\langle h ,\delta \rangle + \frac{\ii \langle \gamma_N , \gamma_S \rangle}{r_{NS}}
\right)  (\cos \theta_N-1 )\dd \phi
\; \simeq \; \frac{\widetilde \beta}{4\pi} (1-\cos \theta_N )\dd \phi
\,,  
\\
\mathcal{A}_d^0 &\; \simeq \; 
- \frac{1}{2} (1+ \cos \theta_N) \dd \phi 
+  \ii \delta^0 
(1-\cos \theta_N) \dd \phi 
,
\\
\Sigma (H)|_{r_N}  &\; \simeq \; \frac{1}{r_N} \frac{\widetilde \beta}{4\pi}  \,,
\qquad 
\frac{\partial \Sigma}{\partial H_0} |_{r_N} \simeq \frac{\ii \gamma_N^0}{r_N} \, ,
\label{eq:near_the_pole_behavior_Sigma}
\end{align}
where $(r_N, \theta_N,\phi)$ are spherical coordinates centered at the north pole, and the last expression can be derived from attractor equations combined with $4d$ regularity conditions \cite{Boruch:2023gfn}.
Defining new coordinates,
\be 
r_N \= 
\frac{4\pi}{\widetilde \beta} 
\frac{\rho_N^2}{4} ,
\qquad 
\hat{\tau} \= \frac{2\pi}{\widetilde \beta} t ,
\qquad \hat{\tau} \; \sim \; \hat{\tau} + 2\pi ,
\ee
and switching to the corotating coordinate $\phi_{\text{cor}}$, the $4d$ metric simplifies to
\be 
\dd s^2_{4d} \; \simeq \; \dd \rho_N^2 + \rho_N^2 \,
\biggl(  
d \Bigl( \, \frac{\theta}{2} \, \Bigr)^2 + 
\cos^2 \Bigl( \,\frac{\theta}{2} \,\Bigr) \,
\dd \hat{\tau}^2 +
\sin^2 \Bigl( \,\frac{\theta}{2} \,\Bigr) \,
\dd \phi_{\text{cor}}^2 
\biggr)
\= 
\dd s^2_{\mathbb{R}^4}
\,,
\ee
as~$\rho_N \to 0$,
which is exactly the Euclidean flat space metric on $\mathbb{R}^4$.%

The gauge field $A^0$ can be conveniently rewritten in terms of functions \eqref{eq:L_function_Shmakova}, \eqref{eq:Q_function_Shmakova} and near the poles takes the form 
\be 
A^0 |_{r_N} \= \ii \frac{(H^0)^2 L}{\Sigma(H)^2} (\dd t + \omega_E) + \mathcal{A}_d^0 \, \, |_{r_N}
\,
\; \simeq \; 
- \frac{4 \pi \gamma^0_N}{\widetilde \beta} (\dd t + \omega_E) 
+ \mathcal{A}_d^0 \,\,|_{r_N}
,
\ee
where we used that close to the pole
\be 
L(H)|_{r_N} \= \frac{\Sigma(H)}{(H^0)^2} 
\frac{\partial \Sigma}{\partial H_0}
|_{r_N} \;\simeq \; \frac{\ii \widetilde \beta}{4\pi \gamma^0_N} \,,
\label{eq:L_near_the_pole}
\ee
which can be derived using the expressions from Appendix~\ref{app:explicit_dipole_charges}. 
Going now to corotating coordinates $\phi \to \phi_{\text{cor}}$ we have that
\be 
\dd t + \omega |_{r_N}\; \simeq  \;
 \frac{1+\cos \theta_N}{2}\dd t 
 + \frac{\widetilde \beta}{2\pi}  \frac{1-\cos \theta_N}{2} 
 \dd \phi_{\text{cor}} ,
\ee
and we can write the gauge field near the poles as simply
\be 
A^0 |_{r_N} \; \simeq \; - \frac{4\pi \ii \delta^0}{\widetilde \beta} \dd t
- \dd \phi_{\text{cor}} .
\ee
Going further to corotating $\psi$-coordinates $\psi \to \psi_{\text{cor}}$ the first term cancels, and we arrive at a simple expression
\be 
\dd \psi + A^0 |_{r_N} \; \simeq  \; \dd \psi_{\text{cor}} - \dd \phi_{\text{cor}} .
\ee
Lastly, we note that the CY volume factor given by \eqref{eq:CY_volume_space_dependent} goes to a constant near the pole. Because of the behavior of the entropy function near the poles \eqref{eq:near_the_pole_behavior_Sigma}, we simply need to argue that the function $Q$ goes to a constant. This can be seen from the relation with the entropy function as follows 
\be 
Q^3 |_{r_N} \= \frac{\Sigma^2+L^2 (H^0)^2}{H^0} |_{r_N} 
\; \simeq \; \frac{(\widetilde \beta)^2}{(4\pi)^2 \gamma^0_N r_N} 
+ L^2 |_{r_N} \frac{\gamma^0_N}{r_N} 
\; =  \; O(1) \; \equiv \; Q_N^3 ,
\ee
where we again used \eqref{eq:L_near_the_pole}. Therefore, we  find that close to the pole, the volume goes to a constant, which we denote as $\widetilde{V}_{\IIA,N}$, and we obtain
\begin{align}
\dd s_{5d}^2 \; \simeq \;
(2\widetilde{V}_{\IIA,N})^{2/3} (\dd \psi_{\text{cor}} - \dd \phi_{\text{cor}})^2 
+(2\widetilde{V}_{\IIA,N})^{-1/3} \dd s^2_{\mathbb{R}^4} 
.
\end{align}
The metric thus takes the form of flat space $\mathbb{R}^4$ trivially fibered on $S^1_\psi$, twisted in the $\phi_{\text{cor}}$ direction as one goes around the $S^1_\psi$. Therefore, the metric near the poles is manifestly smooth. 

\subsection{Contribution to the flat space gravitational index}

Let us now discuss the $5d$ interpretation of the uplifted saddle \eqref{eq:5d_uplifted_BH_metric_Start}--\eqref{eq:5d_uplifted_BH_metric_End}. 
The solution describes a smooth $5d$ Euclidean black hole with flat $S^1_{\widetilde{\beta}} \times \mathbb{R}^4$ asymptotics for~$\Gamma^0=1$. The $5d$ black hole rotates in both $\psi$ and $\phi$ directions and can be viewed as inducing $4d$ black hole charges through a $5d\to 4d$ reduction \cite{Gaiotto:2005gf},
\be 
(\Gamma^0 = 1 , \Gamma^A , \Gamma_A, \Gamma_0) \= 
\left(1 , \Gamma^A , 
\Gamma_A^{(5d)} + \frac{D_{ABC}\Gamma^B \Gamma^C}{2} , 
-\frac{D_{ABC}\Gamma^A\Gamma^B\Gamma^C}{6} 
- \Gamma^A \Gamma_A^{(5d)} - 2J_L
\right) \,.
\ee
The $5d$ black hole additionally rotates along the $\phi$ direction with angular velocity 
$\Omega_R^{5d} = \dfrac{2\pi \ii}{\beta_{5d}}$.
In other words, the rescaled solution obeys 
\be 
\widetilde \beta \, \Omega_\phi
\= \beta_{5d} \, \Omega_{R}^{5d} \= 2\pi \ii \, .
\ee
The scalars in the $5d$ geometry are finite and take asymptotic values $Y^A_\infty$. We can, therefore, interpret this solution as a saddle of the grand-canonical partition function computed in the Gibbons-Hawking prescription 
\be 
\mathcal{I}
\= 
\Tr
(e^{2\pi \ii J_R} e^{-\beta_{5d}(H-E_{BPS})} ) \, .
\ee
After identifying $e^{2\pi \ii J_R} = (-1)^F$ this becomes the $5d$ gravitational index. In this way, the construction relates $5d$ and $4d$ gravitational indices at the level of their leading semiclassical saddles.

As an example, let us study the contribution of the leading saddle in pure $5d$ supergravity. In $4d$, this corresponds to having a single vector multiplet $\nv = 1$. 
Naturally, the answer is simply given by the value of $S_{\text{ext}}$ for the corresponding $4d$ black hole. Evaluated in terms of the $5d$ charges
\be 
(\Gamma^0, \Gamma^1, \Gamma_1, \Gamma_0) \= (1, 0, -\frac{3^{\frac{2}3} d_1^{\frac{1}3}}{2} Q_\Gamma, -2J_L)
\,,
\ee
(where we denote $d_1 = D_{111}$) it takes the form 
\be 
\log \mathcal{I} \= \sqrt{-\frac{8(\Gamma_1)^3}{9d_1} - (\Gamma_0)^2 } \=  \pi \sqrt{Q_\Gamma^3 - 4 J_L^2}
\,.
\label{eq:Example1_5d_black_hole_free_energy}
\ee
This connection is well-known for extremal black holes \cite{Gaiotto:2005gf,Gaiotto:2005xt}. The novelty here is that we obtain it from a gravitational path-integral computation of the index, with \eqref{eq:Example1_5d_black_hole_free_energy} appearing as the
(appropriately subtracted) free-energy of a finite-temperature black hole saddle that contributes to the gravitational index.

Our construction above  
generalizes some of the solutions recently discussed in the literature. 
In particular, Anupam, Chowdhury and Sen \cite{Anupam:2023yns} constructed a finite-temperature saddle that contributes to the gravitational index of supergravity in $5d$ asymptotically flat space for~$\nv = 3$, by taking a supersymmetric limit of the Cvetic-Youm solution \cite{Cvetic:1996xz} keeping the temperature finite. 
The solutions presented in this section, obtained from the Bates-Denef class which are directly supersymmetric, are valid for an arbitrary number of vector multiplets~$\nv$. Therefore they provide index saddles for more general compactifications. 
In  Appendix \ref{app:match_with_sen} we explicitly show in a simple example that our construction reproduces the results of \cite{Anupam:2023yns}.

\section{The gravitational index of $5d$ black strings}
\label{sec:index-5d-black-strings}

We now turn to the solutions with vanishing total D6 charge, $\Gamma^0 = 0$. 
At extremality, this is the case in which a $4d$ black hole is uplifted to a black string, wrapped 
along the M-theory circle with horizon topology~$S^2 \times S^1_M$.
The reason why the horizon topology is different compared to the $5d$ black hole discussed in the previous section is that 
the D6 charge plays the role of the NUT charge from the perspective of the $\psi$-circle, 
making it contractible in the bulk on the black hole horizon. 
This leads to a non-trivial Hopf fibration of $S^2 \times S^1_M$, making the horizon become $S^3$in $5d$. When  $\Gamma^0 = 0$ the Hopf fibration is trivial, and the horizon topology is $S^2 \times S^1_M$, corresponding to a black string wrapping the M-theory circle, which remains compact.

As in the previous section, we begin by describing how to properly take the decompactification limit with 
the new boundary conditions in a way that allows us to keep the five-dimensional temperature finite. 
As will become clear below, the limit of the moduli needs to be taken in a slightly different way in this case, 
which allows us to stay at a finite distance from the near-horizon region. 
The resulting saddle then contributes to the gravitational index of $5d$ supergravity in asymptotically flat space. 

We then take a further decoupling limit of the near horizon AdS$_3 \times S^2$ region by going to zero 
temperature in the $5d$ solution while simultaneously zooming in onto the 
near-horizon region.\footnote{This is  similar to the near-horizon limit taken by Sen for supersymmetric black holes in~\cite{Sen:2008yk}, 
where the result is an AdS$_2$ (Jackiw-Teitelboim) black hole.} 
We show that the resulting geometry is a saddle whose contribution to the partition function matches the 
elliptic genus of MSW CFT \cite{Maldacena:1997de} at leading order in the central charge. 
The geometry of this saddle point takes the form of a spinning BTZ black hole times a 2-sphere with 
periodic boundary conditions for the fermions around the time circle.

\subsection{Uplifting the index saddle: from the $4d$ black hole to the $5d$ black string}
We start with a straightforward uplift of the index saddle of the $4d$ black hole to that of a $5d$ black string 
wrapped along the M-theory circle. 
The metric is simply described by \eqref{eq:Euclidean-metric-uplift} in terms of the harmonic functions
\begin{align}
H^0 &\=  h^0 +\ii \delta^0 \left( 
\frac{1}{r_N} - \frac{1}{r_S}
\right), \\ 
H^A &\= h^A + \frac{\Gamma^A}{2} \left( 
\frac{1}{r_N} + \frac{1}{r_S}
\right)   
+\ii \delta^A \left( 
\frac{1}{r_N} - \frac{1}{r_S}
\right) , \\ 
H_A &\= h_A 
+\ii \delta_A \left( 
\frac{1}{r_N} - \frac{1}{r_S}
\right) ,\\ 
H_0 &\= h_0 + \frac{\Gamma_0}{2} \left( 
\frac{1}{r_N} + \frac{1}{r_S}
\right)   
+\ii \delta_0 \left( 
\frac{1}{r_N} - \frac{1}{r_S}
\right) .
\end{align}
As we have not taken any decompactification limits, 
the proper lengths of the $\psi$-circle and the time circle are finite in $5d$ units, 
and correspondingly the metric is asymptotically $S^1_\beta  \times S^1_M \times \mathbb{R}^3$. 

\paragraph{Topology and smoothness:} 
As before, to see the topology of the horizon, we enclose the horizon with a surface that is infinitesimally close to it at every point. 
In ellipsoidal coordinates, we can describe it simply by taking a constant $\mu$ surface $\mu = \mu_0$. 
We then imagine blowing up this surface up to infinity $\mu_0 \rightarrow \infty$, where the induced metric is
\be 
\dd s^2_\infty \= (2\widetilde V_{\IIA, \infty})^{2/3} \, \dd \psi^2 
+ \frac{e^{2\mu_0}}{(2\widetilde V_{\IIA, \infty})^{1/3}}\, \dd \Omega_2^2 \,,
\ee
from which we see that it is topologically $S^1 \times S^2$. 
Since in taking the parameter $\mu_0 \to \infty$ we have not crossed any poles, 
we see that the horizon has the same topology.

In general, to ensure the smoothness of the uplifted metric, we want to make sure that any possible 
Dirac strings introduced by the one-form fields $\omega_E$ and $A^0$ are merely coordinate singularities. 
For the field $\omega_E$, this is ensured by the regularity condition~\eqref{eq:north_south_distance}
imposed on the four-dimensional geometry. 
For the field $A^0$, the analysis is in the five-dimensional geometry and we perform it in Appendix~\ref{app:Dirac_strings_A0}. 
Because now we have a vanishing D6 charge, we find that there are no additional Dirac-Misner strings 
introduced in the geometry and, consequently, the metric is smooth on the horizon for any value 
of the dipole charge $\delta^0$.

As in the previous section, we now interpret the resulting solution from the perspective of the $5d$ gravitational path integral. The solution is a smooth, finite-temperature $5d$ black hole with 
$S^1_\beta  \times S^1_M \times  \mathbb{R}^3$ asymptotics. 
The black hole has an $S^1 \times S^2$ horizon, carries $5d$ electric charges,  and importantly rotates in the $\phi$ direction with angular velocity 
\be 
 \beta_{5d} \, \Omega_R^{5d} \= 2\pi \ii \,. 
\ee
Therefore, we interpret the resulting solution as a contribution to the $5d$ gravitational index 
\be 
\mathcal{I}_{\text{grav}}^{\mathbb{R}^3 \times S^1_M \times S^1_\beta} 
\=  
\Tr
\left( (-1)^F e^{-\beta_{5d} \left(H - E_\text{BPS}(\Gamma_{5d},\,J_L)\right)} \right) ,
\ee
evaluated with asymptotic $\mathbb{R}^3 \times S^1_M \times S^1_\beta$ boundary conditions and the identification~$e^{2\pi \ii J_R} = (-1)^F$.  
In particular, the free energy of the uplifted D4-D0 solution takes the value 
\be 
\log \mathcal{I}_{\text{grav}}^{\mathbb{R}^3 \times S^1_M \times S^1_\beta}   \= 2 \pi \sqrt{ \Gamma_0 \frac{D_{ABC}\Gamma^A \Gamma^B \Gamma^C}{6}} \,.
\label{eq:gravitation-index-black-string-on-shell}
\ee
We will explain below that this result exactly matches the leading contribution to the elliptic genus of MSW CFT$_2$.

\subsection{Taking the decoupling limit at ``finite temperature"}

We now take a further decoupling limit, which allows us to isolate a saddle with AdS$_3 \times S^2$ asymptotics and finite temperature on the AdS$_3$ boundary, corresponding to the index of the BTZ black hole. 

For simplicity, we will focus on the D4-D0 case, in which the total D2 monopole charge is vanishing, i.e.,~$\Gamma_A=0$. 
To obtain this new decoupling limit, we take the extremal limit of the $5d$ black hole saddle while simultaneously zooming in on the near-horizon region. 
More precisely, we again take the proper length of the M-theory circle to infinity as in \eqref{eq:relation_R_M_and_moduli} and \eqref{eq:parameters_scaling_Lambda}.

Let us reanalyze the behavior of the constant~$h$ with respect to the four-dimensional moduli~$j^A$.
Focusing only on the nonvanishing terms at large $\Lambda$, we have 
\be 
h \=
\left( O(\Lambda^{-5}) \,, \, O(\Lambda^{-1}) \,, \, O(\Lambda^{-1}) \,, \,
\frac{\sqrt{j^3}}{2\sqrt{3}} = O(\Lambda^3)
\right).
\ee
This implies that, in contrast with the solutions of the previous section, when we take large $\Lambda \gg 1$ the interesting part of the geometry localizes onto small coordinate regions where $|\xvec - \xvec_{N/S}| \sim 1/\Lambda^3$. To zoom in onto the near horizon region, we therefore rescale the coordinates as $\xvec \rightarrow \xvec/\Lambda^3$, $t \rightarrow t\Lambda^3$, which leads the harmonic function to behave as 
\be 
H(h,\xvec) 
\= h + \frac{\gamma_N}{r_N} + \frac{\gamma_S}{r_S} 
\= \Lambda^3 
\left( \frac{h}{\Lambda^3} + 
\frac{\gamma_N}{\Lambda^3 r_N} + 
\frac{\gamma_S}{\Lambda^3 r_S}
\right) \= \Lambda^3 H(h/\Lambda^3 ,\, \xvec/\Lambda^3) \,.
\ee
This allows us to take the limit $\Lambda \rightarrow \infty$ in a simple way. 
The new metric is given by \eqref{eq:5d_uplift_formula_1}, with 
\be 
h \=
\left( 0 , 0, 0 , 
\frac{1}{\Lambda^3} \frac{\sqrt{j^3}}{2\sqrt{3}}
\right) 
\; \equiv \;
\left( 0 , 0, 0 , 
h_0
\right) .
\ee
In the new coordinates, the distance between the north and south poles \eqref{eq:north_south_distance} remains finite and is now written in terms of the rescaled periodicity of the new time coordinate,\footnote{Note that with this rescaling the dimensionless parameter $\widetilde \beta$ is different than the one we defined in \eqref{eq:definition-tilde-beta-1} for the decompactification limit of black holes, rather than black strings. }  
\be 
\widetilde \beta \; \equiv \;  
\frac{\beta}{\ell_4}
\frac{1}{\Lambda^3} ,
\ee
which becomes a finite periodicity of the time coordinate in the near-horizon region.
Importantly, even though $\widetilde \beta$ stays finite, we see that in $\Lambda \to \infty$ limit the $5d$ proper length of the Euclidean time circle goes to infinity
\be 
\frac{\beta_{5d}}{\ell_5} \;\sim \;\Lambda^2 \;\rightarrow \; \infty \,,
\ee
in contrast to solutions in the previous section. Therefore, the near-horizon geometry effectively decouples from the asymptotic part of the $5d$ geometry.
As we have taken the extremal limit from the~$5d$ asymptotic perspective, the attractor mechanism ensures 
that the~$5d$ scalars are constant in the resulting geometry and are fixed in terms of the monopole D4 charges
\be 
Y^A \= \frac{\Im t^A}{\widetilde{V}_{\IIA}^{1/3}} \= 
2^{1/3} \frac{y^A(H)}{Q^{1/2}} 
\= 2^{1/3} \frac{\Gamma^A}{U} 
\,, \qquad 
U\;\equiv \; \left(\frac{D_{ABC}\Gamma^A \Gamma^B \Gamma^C}{3} \right)^{1/3}\,.
\ee 

To understand the resulting geometry better, we will use ellipsoidal coordinates centered at the north pole and the south pole
\begin{align}
x &\= \frac{r_{NS}}{2} \sinh{\mu} \sin \theta \cos \phi\,, 
\qquad 
y \= - \frac{r_{NS}}{2} \sinh{\mu} \sin \theta \sin \phi
\,, 
\qquad
z \= \frac{r_{NS}}{2}\cosh{\mu} \cos \theta \,, 
\end{align}
\begin{align}
\xvec_N &\= \left(0,0, \frac{r_{NS}}{2} \right) , 
\qquad 
\xvec_S \=  \left(0,0, -\frac{r_{NS}}{2} \right).
\end{align}
In these coordinates, the harmonic functions are
\begin{align}
\begin{split}
\label{eq:harmonic_funcs_zeroD6}
    H^0 \= \frac{4}{r_{NS}} \frac{\ii \delta^0 \cos \theta}{\cosh^2 \mu - \cos^2 \theta} 
\,, \qquad & 
H^A \= \frac{4}{r_{NS}} \frac{\frac{1}{2}\Gamma^ A \cosh \mu}{\cosh^2 \mu - \cos^2 \theta} 
\,, \\
H_A \=  
\frac{4}{r_{NS}} \frac{\ii \delta_A \cos \theta}{\cosh^2 \mu - \cos^2 \theta} 
\,, \qquad &
H_0 \= h_0 + 
\frac{4}{r_{NS}} \frac{\frac{1}{2}\Gamma_0 \cosh \mu}{\cosh^2 \mu - \cos^2 \theta} 
\,.
\end{split}
\end{align}
Using formulas~\eqref{eq:explicit_dipole_charges_noD6},  
we obtain simple expressions for the dipole charges, 
\begin{align}
\ii \delta^0 
\= - \frac{\ii}{2\sqrt{2}} \sqrt{\frac{U^3}{\Gamma_0}}
\,, \qquad
\ii \delta^A \= 0
\,, \qquad
\ii \delta_A 
\= \frac{\ii}{2\sqrt{2}} \sqrt{\frac{\Gamma_0}{U^3}} D_{ABC} \Gamma^B \Gamma^C 
\,, \qquad
\ii \delta_0 \= 0
\,, 
\end{align}
and the entropy function is $\Sigma(\Gamma) = \sqrt{2 \Gamma_0 U^3}$. 
The metric can now be written as 
\begin{align}
\dd s^2_{5d} &\= 
\frac{1}{Q^2} \left(
\dd t + \omega_E 
+ \ii L(\dd \psi + \mathcal{A}_d^0) \right)^2 
+Q \dd s^2_{\text{TN}} \,,  
\\
\dd s^2_{\text{TN}}
&\=
\frac{1}{H^0} 
(\dd \psi + \mathcal{A}_d^0)^2 
+ H^0 \dd x^m \dd x^m 
\,,
\end{align}
where the explicit form of the functions is
\begin{align}
\omega_E &\= 
\frac{4 \sin^2 \theta \dd \phi}{\cosh(2\mu) - \cos(2\theta)} 
\left( 
\langle h,\delta \rangle \cosh \mu 
+ \frac{\ii \langle \gamma_N, \gamma_S \rangle}{r_{NS}}
\right)
,\\
\mathcal{A}_d^0 &\= 
\ii \delta^0 \cosh \mu
\frac{4 \sin^2 \theta \dd \phi}{\cosh(2\mu) - \cos(2\theta)} 
\,,\qquad
Q\= \ii \frac{2\sqrt{2} \sqrt{\Gamma_0 U}}{r_{NS}\cos \theta}
\,,
\qquad
L\= 
- h_0 - \frac{4 \Gamma_0 \cosh \mu}{r_{NS}\cos^2 \theta}
\,,
\end{align}
and $H^0$ now takes the form for the multi-TaubNUT metric with purely imaginary dipole charges, as given in \eqref{eq:harmonic_funcs_zeroD6}.
As before, the new black hole geometry rotates now along both the $\phi$ and $\psi$ directions with angular velocities 
\be 
\Omega_\phi \; \equiv \; \Omega_R \= \frac{2\pi \ii}{\widetilde \beta} \,, 
\qquad 
\Omega_\psi \; \equiv \; \Omega_L \= \frac{4 \pi \delta^0}{\widetilde \beta} \,.
\ee

To write the metric in a more familiar form, 
we perform the coordinate transformation 
\be 
t \= 
\frac{\widetilde \beta}{2\pi} \tads
-
\frac{\widetilde \beta - 8\pi \langle h,\delta \rangle}{2\pi} \ii  \sigma \,, 
\qquad 
\psi \= 
2 \delta^0 (-\sigma + \ii \tads
) , \qquad \mu = 2\xi \,,
\label{eq:change_to_BTZ_coordinates}
\ee
which leads to 
\be 
\dd s^2 \= 4U^2 (\sinh^2 \xi \dd \tads^2 +\dd \xi^2
+\cosh^2 \xi \dd \sigma^2
) 
+ U^2 (\dd \theta^2 + \sin^2 \theta 
(\dd \phi +(\dd \tads 
-
\ii \dd \sigma)  )^2)
\,.
\ee
The coordinate identifications in this metric follow from those in the $5d$ geometry,
\be 
(t, \psi, \phi) 
\; \sim  \; 
(t +\widetilde \beta ,\psi + \ii \widetilde \beta \Omega_\psi
,\phi + \ii\widetilde \beta\Omega_\phi) 
\; \sim \; (t,\psi + 4\pi , \phi)  
\; \sim \; (t,\psi , \phi + 2\pi)  
\,.
\ee
The first identification corresponds to the contractible cycle at the horizon in the bulk, which is a rotating BTZ black hole. The other two are conventional identifications for the $\psi$ and $\phi$ rotation axes.
 
If we further transform
\be 
\phi' \= \phi + \tads
-
\ii \sigma
\,,
\ee
then the metric takes locally a product form,
\be \label{eq:localBTZ}
\dd s^2 \= 4U^2 (\sinh^2 \xi \dd \tads^2 +\dd \xi^2
+\cosh^2 \xi \dd \sigma^2
) 
+ U^2 (\dd \theta^2 + \sin^2 \theta 
(\dd \phi' )^2)
\,,
\ee
with the twisted identifications 
\be \label{eq:twistedid}
(\tads , \sigma ,\phi') \;\sim \;
(\tads + 2\pi, \sigma , \phi') \; \sim \; 
(\tads + 2\pi \tau_1, \sigma + 2\pi \tau_2 , \phi' + 2\pi(
\tau_1 
-
\ii \tau_2)) \sim
(\tads , \sigma , \phi' + 2\pi) \,,
\ee
with 
\be \label{eq:torusmoduli}
\tau_1 
\= -\ii  \frac{1}{2\delta^0} 
\frac{\widetilde \beta - 8\pi \langle h ,\delta \rangle}{\widetilde \beta - 4\pi \langle h ,\delta \rangle}
, \qquad
\tau_2 
\= 
-\frac{1}{2\delta^0} \frac{\widetilde \beta}{\widetilde \beta - 4\pi \langle 
h , \delta  \rangle} . 
\ee
The equations~\eqref{eq:twistedid} and \eqref{eq:torusmoduli}
also determine the identifications on the boundary torus of AdS$_3$.\footnote{Here we have denoted the modular parameter by~$\tau=\tau_1+\ii \tau_2$ in accord with the AdS$_3$ frame. The natural modular parameter in the BTZ frame would be~$-1/\tau$, which is, of course, related by a modular transformation.}

\subsection{Relation to the index of the MSW CFT$_2$}

The relation between the Euclidean BTZ black hole and the thermal AdS$_3$ geometry is well-known. 
Both configurations correspond to the same hyperbolic manifold with a boundary torus, namely~$H_3^+/\mathbb{Z}$, 
the difference being the identification of the coordinates that correspond to space and time in the Lorentzian geometry. 
In the BTZ black hole, the time circle is contractible, while in thermal AdS$_3$, the space circle is contractible. 
The modular parameters of the respective boundary tori are related by a modular transformation 
as~$\tau_{BTZ} = -1/\tau_{AdS_3}$. 
This relation (and its extension to the infinite number of~$SL(2,\mathbb{Z})$ images~\cite{Maldacena:1997ih}) has been used 
in the calculation of the complete partition function of the bulk in the AdS$_3$/CFT$_2$ correspondence~\cite{Dijkgraaf:2000fq},
and for the microcanonical AdS$_2$ version in~\cite{Murthy:2009dq}.

Although the motivation for the Farey-tail expansion came from a supersymmetric index, 
namely the elliptic genus~\cite{Dijkgraaf:2000fq,deBoer:2006vg}, the precise supersymmetric 
finite-temperature BTZ black hole geometry dual to the elliptic genus 
remained unclear. 
Our construction in the previous subsection of the decoupled non-extremal supersymmetric 
BTZ black hole fills this gap, as we now explain.

Firstly, we note that the extension of the near-horizon $3d$ (times $S^2$) geometry to the asymptotically flat $5d$ geometry is \emph{different} for AdS$_3$ and the BTZ black hole. 
Indeed, the embedding into the asymptotically flat five-dimensional geometry 
chooses a notion of time on the boundary AdS$_3$ torus. From \eqref{eq:change_to_BTZ_coordinates}, we see that the natural notion of time chosen by the $5d$ geometry is the one that leads to BTZ.

The embedding of~AdS$_3$ was already known in the context of supertubes~\cite{Maldacena:2000dr,deBoer:2008fk}. 
This geometry is a supersymmetric global AdS$_3$ (with~$S^2$ fibered over it) and was described in~\cite{Maldacena:2000dr, deBoer:2008fk} in the Lorentzian theory. 
If we consider its Wick rotation and periodically identify the time coordinate,  
we obtain a supersymmetric thermal AdS$_3 \times S^2$. From a five-dimensional perspective, this geometry can be obtained as a region of uplifted four-dimensional D6-$\bar{\text{D6}}$ bound state carrying a $U(1)$ flux~\cite{deBoer:2008fk}. 
In the Euclidean continuation of these solutions, the AdS$_3 \times S^2$ is embedded in the five-dimensional geometry 
with a notion of time consistent with the thermal AdS$_3$ interpretation.

From a different viewpoint, the supersymmetric AdS$_3 \times S^2$ geometry whose conformal boundary is a torus was studied directly in the Euclidean version of $5d$ supergravity  in~\cite{Ciceri:2023mjl}. Here, the motivation was to construct bulk supersymmetric 
configurations dual to the elliptic genus of the boundary SCFT$_2$. 
The method of construction was to begin with local AdS$_3 \times S^2$ geometry and find a twist that admits Killing spinors in the $5d$ supergravity.

One can check that the geometry obtained by either of these two constructions is precisely related to the one discussed in the previous subsection by a modular transformation. 
The map between the BTZ coordinates of \eqref{eq:localBTZ} and the thermal AdS$_3$ coordinates of~\cite{Ciceri:2023mjl} is 
$(t_{\text{BTZ}},\xi , \sigma)_{\text{here}} = (\psi , \rho , t_E)_{\text{there}}$.
Similarly, the modular parameters are related by a modular transformation. 

We now review the argument of~\cite{Ciceri:2023mjl} as to how these solutions contribute to the elliptic genus of 
the MSW CFT$_2$ of the black string.
We use the coordinates and modular parameter corresponding to the thermal AdS$_3$, in accordance with the usual conventions.

The elliptic genus of an~$\mathcal{N}=(0,4)$ SCFT$_2$ is defined as\footnote{One 
can include other~$U(1)$ charges in the left-moving sector since they automatically commute 
with the right-moving supercharge. In a~$(4,4)$ CFT$_2$ there is always 
one such charge, namely the left-moving R-charge. For the BMPV black hole, this would be $J_L$. 
In supergravity, there is always at least one conserved charge from the bulk graviphoton.} 
\be 
\chi(\tau) \=  \Tr_{\text{NS}} (-1)^F q^{L_0} \, \bar{q}^{\bar{L}_0- \bar{J}_3},
\qquad
q=e^{2 \pi \ii \tau}, 
\qquad
\tau=\tau_1+\ii\tau_2 ,
\ee
where $L_n$, $\bar{L}_n$ denote the Virasoro generators of the left-moving and right-moving sectors, respectively, and $\bar{J}_3$ is the Cartan of the right R-symmetry $SU(2)_R$ corresponding to the $S^2$ rotation group. We now want to identify from this the appropriate boundary conditions for the path integration over boundary fields. For this, we set $(-1)^F=e^{2\pi \ii \bar{J}_3}$ and rewrite the elliptic genus as 
\begin{align}
\begin{split}
    \Tr_{\text{NS}} \, (-1)^F q^{L_0} \, \bar{q}^{\bar{L}_0- \bar{J}_3} &\=  
\Tr_{\text{NS}} \, e^{2\pi \ii \bar{J}_3}  \, 
e^{2\pi \ii (\tau_1 + \ii \tau_2) L_0} \,
e^{-2\pi \ii (\tau_1 - \ii \tau_2)(\bar{L}_0- \bar{J}_3)} 
\\ 
&\= \Tr_{\text{NS}} \, e^{2\pi \ii \tau_1 J} \, e^{-2\pi \tau_2 H} \, e^{2\pi \ii (\tau_1 - \ii \tau_2 + 1) \bar{J}_3} 
\,,
\end{split}
\end{align}
where we introduced the CFT Hamiltonian and angular momentum, $H=L_0 + \bar{L}_0$ and $J=L_0 - \bar{L}_0$, which, as we explained, act as translations in the directions $\sigma$ and $\tau$, respectively.
This form of the trace requires specific boundary conditions for the path integration over boundary fields. In particular, the periodicity conditions for the fermionic fields are 
\begin{align}
    \begin{split}
\Psi \left(\tau, \sigma, \phi'\right) 
&\=  
- e^{2\pi \ii \tau_1 J}\, e^{-2\pi \tau_2 H} \, e^{2\pi \ii (\tau_1 - \ii \tau_2 + 1)\bar{J}_3} \, 
\Psi\left(\tau , \sigma, \phi'\right)
\\ 
&\=  
- e^{2\pi \ii \tau_1 J} \, e^{2\pi \ii (\tau_1 - \ii \tau_2 + 1)\bar{J}_3} \, 
\Psi\left(\tau , \sigma + 2\pi \tau_2, \phi'\right)
\\ 
&\= 
-  
\Psi\left(\tau + 2\pi \tau_1 , \sigma + 2\pi \tau_2, \phi' + 2\pi(\tau_1 - \ii \tau_2 + 1) \right)
\\ 
&\= 
\Psi\left(\tau + 2\pi \tau_1 , \sigma + 2\pi \tau_2, \phi' + 2\pi(\tau_1 - \ii \tau_2) \right)
\,,
    \end{split}
\end{align}
where in the last line we used $\Psi (\tau , \sigma , \phi') = - \Psi (\tau , \sigma , \phi' + 2\pi)$. 
These are exactly the same boundary conditions for the fermionic fields as we obtained from the five-dimensional perspective in~\eqref{eq:twistedid}. 
Thus, we conclude that the BTZ$\times S^2$ saddles obtained in the previous section contribute to the elliptic genus of the MSW CFT.

The contribution of the leading saddle, given by \eqref{eq:gravitation-index-black-string-on-shell}, can be rewritten  in terms of CFT quantities as
\be \label{eq:logIG0cl}
\log \mathcal{I}_\text{grav}^{\mathbb R^3 \times S^1_M \times S^1_\beta} \= 2 \pi \sqrt{\frac{ \Gamma_0 \, c_L}{6}} \,,
\ee
where the left-moving central charge is $c_L=D_{ABC}\Gamma^A \Gamma^B \Gamma^C$. 
This is the well-known result for the logarithm of the elliptic genus of the MSW CFT$_2$ to leading order for large charges $\Gamma^A \gg 1$.
In our solutions the modular parameters $\tau$ and $\overline{\tau}$ take the form 
\be 
\tau 
\= \frac{1}{\ii \delta^0}
\= \ii \sqrt{\frac{24 \Gamma_0}{c_L}}
, \qquad
\bar{\tau} \= \frac{1}{\ii} \frac{4\pi h_0}{\widetilde \beta+4\pi h_0 \delta^0}
,
\ee
and therefore our final answer can be rewritten as
\be 
\log \mathcal{I} \= 
\frac{4\pi \ii \, \Gamma_0}{\tau} \,,
\ee
which is precisely the Cardy formula for a holomorphic modular form. 
The temperature independence of the supersymmetric index in asymptotically $5d$ flat space 
implies the holomorphicity of the elliptic genus in the near-horizon region. This is notable: the holomorphicity 
is not forced by changing the shape of the torus keeping~$\tau$ fixed and taking~$\overline{\tau} \to \infty$;
rather, the boundary torus has a smooth structure with finite modular parameter, and the holomophicity 
is a consequence of supersymmetry in the bulk, thus providing a direct bulk version of the boundary argument.

\section{Discussion}
\label{sec:discussion}

\subsection*{Relation to prior work}

Our new saddles are closely related to the construction of a family of Lorentzian horizonless solutions known as ``bubbling solutions"\footnote{For a detailed review and construction of ``bubbling solutions" see e.g. \cite{Bena:2007kg,Warner:2019jll}.}~\cite{Maldacena:2000dr, deBoer:2008fk,Berglund:2005vb,Bena:2005va,Bena:2006is,Bena:2006kb,Cheng:2006yq}. These solutions can also be constructed by uplifting multicentered extremal Bates-Denef solutions to five dimensions and choosing D-brane charges such that the black hole entropy of individual centers vanishes \cite{Cheng:2006yq}. This can be done by restricting the solutions to be only constructed from D6 and anti-D6 branes, while at the same time turning on worldvolume fluxes in such a way that the total charges match the black hole charges. The centers are connected by noncontractible two-spheres (bubbles) extended along the $\psi$-fibers. To ensure that the are no closed timelike curves, one must impose the integrability condition on each center,
\be 
\langle \gamma_i, H(\xvec_i) \rangle = 0\,.
\label{eq:integrability_bubbles}
\ee
Then, the gauge field $\omega$ does not contain any Dirac-Misner strings in the geometry, and the time direction can be non-compact.
Algebraically, this condition is the same as our regularity condition \eqref{eq:north_south_distance} with $\beta = 0$. 

Our construction differs from the bubbling solutions in that we work in Euclidean time, where time circles are not a pathology, and then we can allow the field $\omega$ to have Dirac-Misner strings in the geometry as long as they are unobservable (i.e., smoothly gauge-removable), which is the case when \eqref{eq:north_south_distance} holds \cite{Hartle:1972ya}. When we do this, the geometry contains two types of bubbles. The first ones are analogous to those in the bubbling solutions, with a non-contractible $S^2$ appearing along the $t$-fibers in between the poles in the base space \cite{Yuille:1987vw}. The second ones appear because the Dirac-Misner strings themselves become now non-contractible $S^2$ bubbles, corresponding to the horizons of the black holes in the geometry. These $S^2$ bubbles are already present in the base space, and they are fibered in the $\psi$ direction to form three-dimensional horizons $S^3$ or $S^2 \times S^1$. Our solutions have a single horizon bubble, but for multiple black holes more bubbles will appear \cite{wip}.

\subsection*{Black rings and other more general solutions}

The construction presented in this paper allows to investigate multiple other saddles. For instance, one can study the $5d$ Euclidean saddles that contribute to the index for the uplift of the multicentered black hole saddles of the $4d$ gravitational index \cite{Denef:2000nb,Bates:2003vx,Denef:2007vg,wip}. 
The uplift of the Lorentzian $4d$ Bates-Denef solutions to $5d$ once again provides inspiration for what to expect for the Euclidean contributions to the index. In this case, the bound state of multiple black holes gets uplifted to spinning black holes at the centers of multi-Taub-NUT geometry together with black rings wrapped on the multi-Taub-NUT background \cite{Gaiotto:2005xt,Elvang:2005sa,Bena:2005ni}, whose radii are determined by the bound state distances between $4d$ black holes. Here, $4d$ black hole centers whose D6-charge is $\Gamma^0_i=1$ or $\Gamma^0_i=0$ correspond in $5d$ to spinning black holes and black rings, respectively. Finally, if setting $\Gamma_i^0 \neq 0$ or $1$ then one finds a black lens solution of \cite{Kunduri:2014kja}; while such solutions do not asymptote to $\mathbb R^{3,1}$ by themselves, the bound states that they form with other black holes can result in a solution that is asymptotically flat.\footnote{This happens when the D6 charges of the other centers are chosen such that the total D6 charge is $1$, i.e., $\sum_i \Gamma_i^0 = 1$.}   
The $4d$ construction of the index saddles that we reviewed in section \ref{sec:new_attractor_saddles_4d_review} allows for the construction of the multi-center black hole saddles that contribute to the $4d$ index. We will present such solutions and a study of their moduli space in upcoming work \cite{wip}.  Applying the uplift to these new saddles should then allow one to thus construct more complicated $5d$ index saddles, like the bound states of spinning black holes with black rings or black lenses.  We hope to also report on these directions in the near future.

\subsection*{Towards general supersymmetric indices in flat space}

The saddles that we have found allow us to smoothly interpolate between the finite-temperature black string index in flat space and the zero-temperature black string index that corresponds to the elliptic genus of the MSW SCFT$_2$, always finding the same value for the index. However, the SCFT index only captures the single black string sector of the gravitational index. More generally, the gravitational flat space index receives additional contributions from saddles that we have not considered, such as the bound multi-center black string and black hole configurations discussed above. We expect that their contribution can be reproduced by taking a product of individual microscopic indices, each corresponding to a different theory determined by the choice of brane charges of each black string.\footnote{See section 5 of \cite{Denef:2007vg} for a discussion about the expected structure of the total flat space index.} The total gravitational index will be a sum over all possible charge assignments for the bound states discussed above. Nevertheless, the match between the gravitational index of the single black string sector and the SCFT$_2$ elliptic genus, represents an initial step in trying to understand the match between flat space finite-temperature gravitational quantities protected by supersymmetry, such as the gravitational index considered in this paper, and microscopic quantities protected by supersymmetry in string theory. While the results above were restricted to solutions that are asymptotically $\mathbb R^5 \times $CY studied at leading order in $G_N$, we expect this limitation to be purely technical. We hope to extend our result for D-brane configurations with different asymptotics and go beyond the semi-classical saddle-point approximation by including quantum corrections subleading in $G_N$ in the near future.

\subsection*{Acknowledgements}

We thank Gustavo Joaquín Turiaci for discussions at the beginning of this project. We also thank David Katona for useful comments. LVI and JB are supported by the DOE Early Career Award DE-SC0025522 and by the DOE QuantISED Award DE-SC0019380. RE is supported by MICINN grant PID2022-136224NB-C22, AGAUR grant 2021 SGR 00872, and State Research Agency of MICINN through the ``Unit of Excellence María de Maeztu 2020-2023'' award to the Institute of Cosmos Sciences (CEX2019-000918-M).
S.M.~acknowledges the support of the J.~Robert Oppenheimer 
Visiting Professorship at the Institute for Advanced Study, Princeton, USA and 
the STFC grants ST/T000759/1,  ST/X000753/1.

\appendix

 \section{Match with Anupam-Chowdhury-Sen}
\label{app:match_with_sen}
In this appendix, we verify in detail that the geometries derived in Section~\ref{sec:uplifted_5d_black_holes_nonzero_D6},  in the case of a single vector multiplet $\nv=1$, generalize the geometries previously found in the literature by Anupam-Chowdhury-Sen (ACS)~\cite{Anupam:2023yns}. 
This match has also been obtained in \cite{Hegde:2023jmp}.

We will denote the components of $H(\xvec)$ as
\be
H(\xvec) = ( H^0 ,  H^1, H_1, H_0) 
\; \equiv\;  (v, p, q, u) \, .
\ee
The algebraic equations determining $y(H)$ can be explicitly solved in this case, and one can write down all the functions entering the metric explicitly (we denote $d_1 \equiv D_{111}$)
\begin{align}
\Sigma(H) &= 
\sqrt{
\frac{1}{3} p^2 q^2 - \frac{8}{9 d_1} q^3 v +
\frac{2 d_1}{3} p^3 u - 2 p q u v - u^2 v^2 
} ,\\
Q(H) &= \frac{d_1 p^2 -2qv}{3^{2/3} d_1^{1/3} v} , 
\qquad 
L(H) = \frac{d_1 p^3 - 3v (pq + u v)}{3v^2}
.
\end{align}
We focus on the monopole charge $\Gamma = (\Gamma^0 , \Gamma^1 , \Gamma_1 , \Gamma_0)$ with 
\be 
\Gamma^0 = v_M = 1 , 
\qquad \Gamma^1 = p_M = 0 , 
\qquad \Gamma_1 = q_M < 0 , 
\qquad \Gamma_0 =u_M > 0 ,
\ee
and correspondingly, the harmonic functions are given by 
\begin{align}
H^0 &= v= \frac{1}{2} \left( 
\frac{1}{r_N} + \frac{1}{r_S}
\right)
+
\ii \delta^0 \left( 
\frac{1}{r_N} - \frac{1}{r_S}
\right)
,
\\ 
H^1 &= p=  
\ii \delta^1 \left( 
\frac{1}{r_N} - \frac{1}{r_S}
\right)
,
\\ 
H_1 &= q= h_q 
+ 
\frac{q_M}{2} \left( 
\frac{1}{r_N} + \frac{1}{r_S}
\right)   
+\ii \delta_1 \left( 
\frac{1}{r_N} - \frac{1}{r_S}
\right)
,
\\ 
H_0 &= u= \frac{u_M}{2} \left( 
\frac{1}{r_N} + \frac{1}{r_S}
\right)   
+\ii \delta_0 \left( 
\frac{1}{r_N} - \frac{1}{r_S}
\right)
,
\end{align}
where only a single component of the constant $h$ remains nonzero as we have already taken the decompactification limit, and we are working with zero D4 charge. 
To match the exact expressions, it will be convenient to write the metric in terms of charge $Q_\Gamma \equiv Q(\Gamma)$ and angular momenta $J_L,\, J_R$. We introduce $Q_\Gamma$ and $J_L$ as
\begin{align}
Q_\Gamma &\equiv Q(\Gamma) = \left( 
\frac{1}{3} d_{1} y^1(\Gamma)
\right)^{2/3} 
= - \frac{2}{3^{2/3} d_1^{1/3}} q_M  , 
\\ 
2 J_L &= \frac{D_{ABC}\Gamma^A \Gamma^B \Gamma^C}{3 (\Gamma^0)^2} 
- \frac{\Gamma^A \Gamma_A}{\Gamma^0}
- \Gamma_0 = 
-\Gamma_0  
= -u_M  
.
\end{align}
The dipole charges in terms written in terms of $Q_\Gamma$ and $J_L$ can be written as 
\be 
\ii \delta = \frac{1}{S_{\text{ext}}} 
\left( 
- \pi \ii J_L , 
 \pi \ii \frac{3^{1/3}Q_\Gamma^2}{2 d_1^{1/3}} ,
- \pi \ii \frac{d_1^{1/3} 3^{2/3}}{2} J_L Q_\Gamma
, 
2\pi \ii \left( 
\frac{Q_\Gamma^3}{8} - J_L^2
\right)
\right)
,
\ee
where $S_{\text{ext}}$ is the extremal entropy given by
\be 
S_{\text{ext}} = \pi \sqrt{Q_\Gamma^3 - 4 J_L^2} .
\ee
We can now explicitly write the right angular momentum as 
\be 
J_R = \frac{\ii}{2} 
\langle h, \delta \rangle r_{12} 
= -\ii 
\frac{3^{1/3}}{4 d_1^{1/3}}
\frac{ h_q Q_\Gamma^2}{ 
\sqrt{Q_\Gamma^3 - 4 J_L^2}} 
r_{NS}
,
\ee
which further allows us to express the rescaled inverse temperature $\widetilde{\beta}$ in terms of $Q_\Gamma, J_L,J_R$ as (note that $h_q<0$)
\be 
\widetilde{\beta} = -\frac{3^{1/3} h_q }{d_1^{1/3}} Q_\Gamma^2 \left( 
 \frac{\pi \ii}{J_R} 
+ \frac{2 \pi }{\sqrt{Q_\Gamma^3 - 4 J_L^2}}
\right) 
.
\ee
Lastly, we need the explicit expressions for gauge fields $\omega_E$ and $\mathcal{A}_d^0$. Once again, introducing ellipsoidal coordinates (here we use an opposite sign for the $\phi$ coordinate compared to the main text)
\begin{align}
x &= \frac{r_{NS}}{2} \sinh{\mu} \sin \theta \cos \phi, 
\qquad 
y = \frac{r_{NS}}{2} \sinh{\mu} \sin \theta \sin \phi
, 
\qquad
z = \frac{r_{NS}}{2}\cosh{\mu} \cos \theta, 
\\ 
\xvec_N &= \left(0,0, \frac{r_{NS}}{2} \right) , 
\qquad 
\xvec_S =  \left(0,0, -\frac{r_{NS}}{2} \right)
,
\end{align}
the gauge fields can be written as 
\begin{align}
\omega_E &= 
- 
\frac{4 \sin^2 \theta \dd \phi}{\cosh(2\mu) - \cos(2\theta)} 
\left( 
\langle h,\delta \rangle \cosh \mu 
+ \frac{\ii \langle \gamma_N, \gamma_S \rangle}{r_{NS}}
\right)
,\\
\mathcal{A}_d^0 &= 
-
\ii \delta^0 \cosh \mu
\frac{4 \sin^2 \theta \dd \phi}{\cosh(2\mu) - \cos(2\theta)} 
, 
\qquad 
\gamma_{N/S} = \frac{\Gamma}{2} \pm \ii \delta .
\end{align}

With this, we are ready to make a comparison with the ACS metric. 
We redefine the angles as 
\be 
\theta_{\text{ACS}} = \frac{\theta}{2} , 
\qquad 
\psi_{\text{ACS}} = \frac{\psi+\phi}{2} , 
\qquad 
\phi_{\text{ACS}} = \frac{\phi-\psi}{2} ,
\ee
and introduce new coordinates 
\be 
t = \frac{-2h_q}{3^{2/3}d_1^{1/3}} t' 
, 
\qquad 
\cosh \mu \equiv 
\frac{3\ii Q_{\Gamma}^2 }{2^{2/3} J_R 
\sqrt{Q_{\Gamma}^3 - 4 J_L^2}} 
\rho^2
.
\ee
To see the match, we need to go to the dimensionless form of the metric in the ACS metric. We then find the following relation is satisfied
\be 
\dd s^2_{\text{us}}(J_L \rightarrow - J_L ; Q_\Gamma \rightarrow 2^{4/3} Q_{(1)}) 
= \frac{\dd s^2_{\text{ACS}} (t \rightarrow \ell_5 t; \rho \rightarrow \ell_5 \rho)}{\ell_5^2},  
\ee
where one needs to use that ACS work in units $\ell_5 = 2^{5/3} $.

\section{Explicit dipole charges}
\label{app:explicit_dipole_charges}
In this appendix we derive the explicit form of the dipole charges introduced in the finite temperature solutions in Sec.~\ref{sec:new_attractor_saddles_4d_review}.
The dipole charges given in \eqref{eq:dipole-charge} can be rewritten as 
\be 
\ii \delta^\alpha = \frac{\ii}{2} I^{\alpha \beta} \frac{\partial \Sigma}{\partial H^\beta} ,
\ee
which in our conventions for the intersection product in the symplectic basis is 
\be 
\ii \delta 
= (\ii \delta^0 , \ii \delta^A , \ii \delta_A , 
\ii \delta_0
)
= \left( 
-\frac{\ii}{2} 
\frac{\partial \Sigma}{\partial H_0}
, 
-\frac{\ii}{2} 
\frac{\partial \Sigma}{\partial H_A}
, 
\frac{\ii}{2} 
\frac{\partial \Sigma}{\partial H^A}
,
\frac{\ii}{2} 
\frac{\partial \Sigma}{\partial H^0}
\right) .
\label{eq:explicit_dipole_charge_eq1}
\ee
Using the explicit solution to the attractor equations by Shmakova, we can write explicit expressions for the derivatives in terms of $L$ and $Q$ (see \eqref{eq:L_function_Shmakova} and \eqref{eq:Q_function_Shmakova}) as 
\begin{align}
\frac{\partial \Sigma}{\partial H_0} &= 
\frac{(H^0)^2 L}{\Sigma(H)} 
, \\
\frac{\partial \Sigma}{\partial H_A} &= 
\frac{ H^0
(L H^A - y^A Q^{3/2})
}{\Sigma(H)}
,\nonumber \\
\frac{\partial \Sigma}{\partial H^A}&=
\frac{
Q^{3/2} D_{ABC} H^B y^C 
-L (D_{ABC} H^B H^C - H^0 H_A)
}{\Sigma(H)}
, \nonumber\\
\frac{\partial \Sigma}{\partial H^0} &= 
\frac{
Q^3 - Q^{3/2}\frac{D_{ABC}H^A H^B y^C}{H^0}
-2 H^0 L^2 
+ \frac{4}{3} L \frac{D_{ABC}H^A H^B H^C}{H^0} 
- 2 L H_A H^A
}{2\Sigma(H)} 
\nonumber
,
\label{eq:explicit_dipole_charge_eq2}
\end{align}
with the entropy function being given by
\be 
\Sigma(H) = \sqrt{Q^3 H^0 - L^2 (H^0)^2} 
.
\ee

For some purposes, it will be useful to have the $H^0 \rightarrow 0$ limit of the above equations. Carefully expanding $Q$ and $L$ (it's useful to switch to $Q_D$ and $L_D$ first) we obtain 
\begin{align}
\frac{\partial \Sigma}{\partial H_0} &= \frac{D_{MNK}H^M H^N H^K}{3 \Sigma(H)}
, \\
\frac{\partial \Sigma}{\partial H_A} &= 
\frac{D_{MNK}H^M H^N H^K}{3 \Sigma(H)} D^{AB} H_B
, \nonumber\\
\frac{\partial \Sigma}{\partial H^A}&=
\frac{
D_{ABC}H^B H^C (2H_0 + H_M D^{MN} H_N) - 
\frac{D_{MNK}H^M H^N H^K}{3} D_{ABC} D^{BG} H_G D^{CF} H_F
}{2\Sigma(H)}
, \nonumber\\
\frac{\partial \Sigma}{\partial H^0} &= 
\frac{
\frac{D_{MNK}H^M H^N H^K}{9} D_{FGH} D^{FJ}H_J D^{GK}H_K D^{HL} H_L
-H_A H^A (2H_0 + H_B D^{BC} H_C)
}{2\Sigma(H)}
,\nonumber
\label{eq:explicit_dipole_charges_noD6}
\end{align} 
where $D^{AB}$ is the inverse matrix of $D_{ABC}H^C$ and $\Sigma$ in the $H^0 \rightarrow 0$ limit takes the form 
\be 
\Sigma (H)= \sqrt{\frac{1}{3}D_{ABC}H^A H^B H^C 
(2H_0 + H_A D^{AB} H_B)
}
.
\ee

\section{Dirac-Misner strings of the gauge field $A^0$}
\label{app:Dirac_strings_A0}
In this section we analyze the regularity of the uplifted metric, 
\be
\dd s_{5d}^2 = 2^{2/3}\widetilde{V}_\IIA^{2/3} \, (\dd\psi+A^0)^2 + 2^{-1/3}\widetilde{V}_\IIA^{-1/3} 
\, \dd s^2_{4d} , \qquad 
\psi \sim \psi + 4\pi
\ee
focusing in particular on possible Dirac-Misner strings introduced by the gauge field 
\be 
A^0 = 
\ii \Phi^0 (\dd t +\omega_E) 
+ \mathcal{A}_d^0 ,
\qquad
\Phi^0 \equiv 
\partial_{H_0} \log \Sigma(H)
,
\ee
where we introduced the chemical potential $\Phi^0$.
We consider a general case with the north pole charge given by 
\be 
\gamma^0_N = \frac{\Gamma^0}{2} + \ii \delta^0 (\Gamma) . 
\ee
To see if there are Dirac-Misner strings in the geometry we evaluate an integral over a closed volume $V$, in the $\mathbb{R}^3$ base space. 
Let us first consider a situation in which $V$ encloses only a north pole and goes to infinity. Upon applying Stokes's theorem twice, we obtain~\cite{Boruch:2023gfn} 
\be 
\int_V \dd\dd A^0  = \int_{S_\infty} A^0_\phi \dd\phi -
\int_{S_N} A^0_\phi \dd\phi  
\ee
where $S_N$ denotes an infinitesimal loop just behind the north pole at $\mu = 0$, and $S_\infty$ denotes an infinitesimal loop at infinity at $\theta = 0$. The $\phi$ component of the gauge field is given by
\be 
A^0_\phi = \ii \Phi^0 \omega_{E,\phi} 
+ \mathcal{A}_{d,\phi}^0 ,
\ee
where in terms of ellipsoidal coordinates we have 
\begin{align}
\omega_E &= 
\frac{4 \sin^2 \theta \dd \phi}{\cosh(2\mu) - \cos(2\theta)} 
\left( 
\frac{\ii \langle \gamma_N , \gamma_S \rangle}{r_{NS}} 
+ \langle h , \delta \rangle \cosh \mu 
\right)
, 
\\ 
\mathcal{A}_d^0 &= 
-\frac{\Gamma^0}{2} \frac{4 \sinh^2 \mu \cos \theta \dd \phi }{\cosh(2\mu) - \cos(2\theta)} 
+ \ii \delta^0 \frac{4 \cosh \mu \sin^2 \theta \dd\phi}{
\cosh(2\mu) - \cos(2\theta)}
.
\end{align}
Using the regularity condition \eqref{eq:north_south_distance}, one finds that these fields take the values
\be 
\omega_{E,\phi}|_{S_\infty} = 0
, \qquad
\omega_{E,\phi}|_{S_N} = \frac{\widetilde{\beta}}{2\pi} \dd \phi
,
\qquad
\mathcal{A}_d^0|_{S_\infty} = - \Gamma^0 \dd \phi 
, \qquad
\mathcal{A}_d^0|_{S_N} = 2\ii \delta^0 \dd \phi
,
\ee
and similarly, the chemical potential at the pole is simply \cite{Boruch:2023gfn}
\be 
\Phi^0 |_{S_N} = \frac{4\pi \ii}{\widetilde{\beta}} \gamma^0_N .
\ee
Putting everything together we find that 
\be 
\int_V \dd\dd A^0 = 4\pi \left( 
\gamma^0_N - \frac{\Gamma^0}{2} - \ii \delta^0 
\right) 
=0 
,
\ee
from which we see that the dipole charges themselves are not subject to any specific quantization condition. 

On the other hand, let us now consider the situation in which $V$ encloses both the north pole and the south pole. In this case, we will obtain 
\be 
\int_V \dd\dd A^0 = 
\int_{S_\infty} A^0_\phi \dd\phi -
\int_{S_{-\infty}} A^0_\phi \dd\phi  ,
\ee
where $S_{-\infty}$ denotes the infinitesimal loop at $\theta = \pi$. Evaluating it explicitly, we now find 
\be 
\int_V \dd\dd A^0 = - 4\pi \Gamma^0 ,
\ee
which indicates the presence of the Dirac-Misner string in the geometry. To remove this string, we need to perform a coordinate transformation in 
\be 
\psi = \psi' + 2\Gamma^0 \phi , 
\ee
which induces a gauge transformation on the gauge field $A^0 \rightarrow A^0 + 2 \Gamma^0 \dd \phi$, thereby removing the Dirac-Misner string from the geometry. For this coordinate transformation to be valid, the periodicities of new coordinates have to be consistent, which implies the quantization of the monopole D6 charge $\Gamma^0$. In particular, to further obtain flat asymptotics this requires 
\be 
\Gamma^0 = 1. 
\ee

\bibliographystyle{utphys2}
{\small \bibliography{Biblio}{}}

\end{document}